\documentclass[lettersize,journal]{IEEEtran}
\usepackage{amsmath,amsfonts}
\usepackage{algorithmic}
\usepackage{algorithm}
\usepackage{array}
\usepackage[caption=false,font=normalsize,labelfont=sf,textfont=sf]{subfig}
\usepackage{textcomp}
\usepackage{stfloats}
\usepackage{url}
\usepackage{verbatim}
\usepackage{graphicx}
\usepackage{cite}
\usepackage{enumitem}

\usepackage{url}
\usepackage{wasysym}

\usepackage{amsmath,graphicx, color}
\usepackage{epstopdf, latexsym, amssymb, cite, ushort, mathtools, caption, url, pgfplots, tikzscale}
\usepackage{booktabs}
\usepackage{ushort}
\usepackage{calrsfs}
\DeclareMathAlphabet{\pazocal}{OMS}{zplm}{m}{n}
\usepackage{makecell}
\usepackage{array}
\usepackage{textcomp}
\usepackage{calc}
\usepackage{tikz}
\usetikzlibrary{shapes,arrows}
\usetikzlibrary{calc}
\usetikzlibrary{positioning}
\usetikzlibrary{arrows}
\usetikzlibrary{arrows.meta}
\usetikzlibrary{arrows,scopes}
\usetikzlibrary{backgrounds}
\usetikzlibrary{patterns}
\pgfplotsset{
    /pgfplots/layers/Bowpark/.define layer set={
        axis background,axis grid,main,axis ticks,axis lines,axis tick labels,
        axis descriptions,axis foreground
    }{/pgfplots/layers/standard},
}
\usepackage{xfrac} 
\usepackage{nicefrac}  
\usepackage{upgreek} 
\usepackage{cases} 
\usepackage{siunitx}
\usepackage{empheq}
\usepackage{xcolor}
\usepackage[font=small]{caption}

\def\BibTeX{{\rm B\kern-.05em{\sc i\kern-.025em b}\kern-.08em
    T\kern-.1667em\lower.7ex\hbox{E}\kern-.125emX}}

\DeclareMathOperator {\sinc}{sinc}

\DeclareMathOperator {\tr}{tr}

\graphicspath{{./fig/}}

\allowdisplaybreaks

\def\h{\mathbf{h}}

\def\0{{\mathbf{0}}}

\def\bPsi{\boldsymbol{\Psi}}
\def\bpsi{\boldsymbol{\psi}}

\def\bLambda{\mathbf{\Lambda}}

\newcommand*{\transp}{{{\mathrm{T}}}}
\newcommand*{\herm}{{{\mathrm{H}}}}

\newlength\fheight
\newlength\fwidth

\begin{document}

\title{{Scalable}-Complexity Steered Response Power\\ based on Low-Rank and Sparse Interpolation}

\author{Thomas Dietzen,~\IEEEmembership{Member,}
Enzo De Sena,~\IEEEmembership{Senior Member,}
Toon van Waterschoot,~\IEEEmembership{Member}

\thanks{
T. Dietzen is with KU Leuven, Dept. of Electrical Engineering (ESAT), Processing Speech and Images, Sint-Katelijne-Waver, Belgium. T. van Waterschoot are with KU Leuven, Dept. of Electrical Engineering (ESAT), STADIUS Center for Dynamical Systems, Signal Processing and Data Analytics, Leuven, Belgium. Enzo De Sena is with University of Surrey, Institute of Sound Recording, Guildford, UK.}
}



\maketitle

\begin{abstract}
The steered response power (SRP) is a popular approach to compute a map of the acoustic scene, typically used for acoustic source localization.
The SRP map is obtained as the frequency-weighted output power of a beamformer steered towards a grid of candidate locations.
Due to the exhaustive search over a fine grid {at all frequency bins}, conventional {frequency domain-based} SRP (conv. FD-SRP) results in a high computational complexity.
{Time domain-based SRP (conv. TD-SRP) implementations reduce computational complexity at the cost of accuracy using the inverse fast Fourier transform (iFFT).}
{In this paper, to enable a more favourable complexity-performance trade-off as compared to conv. FD-SRP and conv. TD-SRP, we consider the problem of constructing a fine SRP map over the entire search space at scalable computational cost. 
We propose two approaches to this problem.}
Expressing the conv. FD-SRP map as a matrix transform of frequency-domain GCCs, we decompose the SRP matrix into a sampling matrix and an interpolation matrix.
While sampling can be implemented by the iFFT, we propose to 
use optimal low-rank or sparse approximations of the interpolation matrix for complexity reduction.
The proposed approaches, refered to as sampling + low-rank interpolation-based SRP (SLRI-SRP) and sampling + sparse interpolation-based SRP (SSPI-SRP), are evaluated in 
{various} localization scenario{s with speech as source signals} and compared to the state-of-the-art. 
The results indicate that SSPI-SRP performs better if large array apertures are used, while SLRI-SRP performs better at small array apertures or a large number of microphones. 
{In comparison to conv. FD-SRP,} two to three orders of magnitude of complexity reduction can achieved, {often times enabling a more favourable complexity-performance trade-off as compared to conv. TD-SRP.} 
A \mbox{MATLAB} implementation is available online.
\end{abstract}


\section{Introduction}

Acoustic source localization \cite{chen2010introduction, brandstein2013microphone, evers2020locata} in noisy and reverberant environments remains an essential task in many multi-microphone signal processing \cite{brandstein2013microphone, gannot2017consolidated, SourceSep2018} applications.
{Classical approaches use explicit models exploiting time differences of arrival (TDOAs). 
This includes approaches based on the generalized cross-correlation (GCC) \cite{knapp1976generalized}, 
subspace-based approaches like MUSIC \cite{Schmidt1986} or ESPRIT \cite{Roy89}, 
sparse estimation approaches like SPICE \cite{Stoica2011}, and steered response power (SRP) mapping \cite{Grinstein2024, dibiase2000high, dibiase2001robust, zotkin2004accelerated, Silverman05, dmochowski2007generalized, do07, do2007fast, tervo2008interpolation, do09, cobos2010modified, marti13, nunes14, lima2014volumetric, velasco2016proposal, salvati2017exploiting, Coteli2018, grondin2019svd, DiazGuerra2021, GarciaBarrios2021, dietzen2021low, Cakmak2022, Grinstein2023}.
In addition, learning-based approaches \cite{grumiaux2022survey, DiazGuerra2021, Grinstein2023} have been proposed recently.}

In SRP, a map of the acoustic scene is computed from the frequency-weighted output power of a delay-and-sum (DSB) beamformer steered towards a set of candidate locations. 
Given this map, the location of a single source may be inferred from its maximum.
The major disadvantage of this {conventional frequency-domain based formulation of SRP (conv. FD-SRP)} is posed by its high computational complexity due to the {fine} grid of candidate locations {evaluated at every frequency bin}. 
It is well known \cite{dibiase2000high, dibiase2001robust, Silverman05, velasco2016proposal} that SRP may equivalently be expressed in terms of time-domain GCCs (TD GCCs) \cite{knapp1976generalized}  of the microphone signals at lags equal to the candidate locations' TDOAs.
{This is exploited in computationally more efficient conventional time-domain SRP (conv. TD-SRP) implementations \cite{dibiase2000high, dibiase2001robust, Silverman05} which make us of the inverse fast Fourier Transform (iFFT). Here, the TDOAs are commonly rounded towards the nearest time-domain sample at the cost of accuracy in particular for small microphone array apertures.}

In the literature, several low-complexity SRP approaches \cite{zotkin2004accelerated, dmochowski2007generalized, do07, do2007fast, do09, cobos2010modified, marti13, nunes14, lima2014volumetric, Coteli2018, grondin2019svd, dietzen2021low} have been proposed. 
The majority of these approaches \cite{zotkin2004accelerated, do07, do2007fast, do09, cobos2010modified, marti13, nunes14, lima2014volumetric, Coteli2018} essentially performs a non-exhaustive search as a strategy to reduce computational cost {of TD-SRP}: In stochastic region contraction \cite{do07, do09} or coarse-to-fine search strategies \cite{zotkin2004accelerated, do2007fast}, the search space is iteratively contracted based on the  SRP map obtained in the previous iteration.  
Volumetric \cite{cobos2010modified, marti13, nunes14, lima2014volumetric} or density-based \cite{Coteli2018} approaches accumulate the SRP function corresponding to the volume surrounding the candidate locations  \cite{cobos2010modified, marti13, nunes14, lima2014volumetric} or a surface element in the spherical harmonics domain \cite{Coteli2018}.
This allows for a coarser spatial grid \cite{cobos2010modified, marti13, nunes14} and (similarly to \cite{zotkin2004accelerated, do07, do09, do2007fast}) enables iterative or hierarchical \cite{marti13, nunes14, Coteli2018} search approaches. 
In \cite{dmochowski2007generalized}, instead of searching over space, the search is performed over TDOAs, at each of which the SRP map is updated for all corresponding candidate locations. 

Recent advances in acoustic mapping and localization using distributed processing \cite{Cakmak2022, Grinstein2023} or deep learning \cite{DiazGuerra2021, Grinstein2023} require a {fine SRP map over the entire search space}, where the non-exhaustive search approaches described above are not immediately applicable.
{In this work, we therefore consider the problem of constructing a fine SRP map at a {scalable} computational cost {to enable a more favourable trade-off between complexity and performance as compared to conv. FD-SRP and conv. TD-SRP.}
This problem has previously been addressed in \cite{grondin2019svd, dietzen2021low}.}
In \cite{grondin2019svd}, a low-rank approximation of the SRP matrix is proposed.
In our previous work \cite{dietzen2021low}, we have introduced a low-complexity SRP approach based on Nyquist-Shannon sampling and interpolation.
{In this approach, the bottleneck for complexity reduction is typically posed by the interpolation operation.}

In this paper, we propose two extensions to our previous work \cite{dietzen2021low}.
{The contributions of this paper to the problem of constructing a fine SRP map at a reduced computational cost are as follows.
\begin{itemize}[leftmargin=*]
\item As compared to other SRP approaches using interpolation \cite{Silverman05, tervo2008interpolation}, we derive scalable, low-complexity interpolation schemes based on global optimality criteria. Expressing the SRP map as a matrix transform of frequency-domain GCCs, we decompose the SRP matrix into a sampling matrix and an interpolation matrix. While sampling can be implemented efficiently by the iFFT, we propose to use low-rank or sparse approximations of the interpolation matrix for further complexity reduction.
These approximations are obtained as the solution to a quadratic optimization problem with low-rank and sparsity constraints, respectively, and are therefore optimal in the squared-error sense.
\item The trade-off between complexity and performance is thoroughly investigated. The proposed approaches, refered to as sampling + low-rank interpolation-based SRP (SLRI-SRP) and sampling + sparse interpolation-based SRP (SSPI-SRP), are evaluated in 
{various} localization scenario{s with speech as source signals} and compared to the low-rank-based SRP approach (LR-SRP) in \cite{grondin2019svd} {and the non-exhaustive spatial search approach in \cite{marti13}}.
The results indicate that among the fine SRP mapping approaches, SSPI-SRP outperforms both SLRI-SRP and LR-SRP over a wide complexity range 
{if a low number of microphones and large array apertures are used, while SLRI-SRP performs better for small array apertures, and both SLRI-SRP and LR-SRP perform better at a larger number of microphones.}
{In comparison to conv. FD-SRP, two to three orders of magnitude of complexity reduction can achieved, often times yielding a more favourable trade-off between complexity and performance in comparison to conv. TD-SRP.} 
\item A \mbox{MATLAB} implementation of the proposed approaches is available online \cite{taslp23Code}.
\end{itemize}}

The paper is organized as follows.
Sec. \ref{sec:SRP} introduces conv. {FD-}SRP and conv. {TD-}SRP.
In Sec. \ref{sec:review}, we review \cite{grondin2019svd} and our previous work \cite{dietzen2021low}.
The proposed low-rank and sparse interpolation approaches are introduced in Sec. \ref{sec:lr_sp_int} and evaluated in Sec. \ref{sec:sim}.
Sec. \ref{sec:conclusion} concludes the paper.

\section{Steered Response Power Mapping}
\label{sec:SRP}

We define the signal and propagation model
in Sec. \ref{sec:sigpropmodel}, revisit SRP in Sec.  \ref{sec:bfperspective}, outline its relation to the TD GCC in Sec. \ref{sec:timedomainGCC}, define a matrix-vector representation in \ref{sec:discreteFreq}, and discuss its complexity in Sec. \ref{sec:compcomp}.
{The notation is defined in Table \ref{tab:notation}.}

\begin{table}[t]
	\caption{{Notation.}}
	\begin{center}
	{
		\begin{tabular}{@{} l p{0.76\linewidth} l } 
			\toprule
			object &  related notation  \\
			\midrule
	boldface letter & upper-case: matrix; lower-case: vector.\\
						matrix $\mathbf{A}$ &  $\mathbf{A}^\transp$: transpose; $\mathbf{A}^\herm$: Hermitian transpose;  $\Re[\mathbf{A}]$: real part; $\tr[\mathbf{A}]$: trace; $ \operatorname{rank}[\mathbf{A}]$: rank; $\operatorname{vec}[\mathbf{A}]$: vectorization; $\Vert\mathbf{A}\Vert_{\mathrm{F}}$: Frobenius norm; $[\mathbf{A}]_{m,m'}$: element at  index $(m,m')$; $ \operatorname{blkdiag}[\mathbf{A}_1, \mathbf{A}_2, \dots ]$: block-diagonal matrix.\\		
						vector $\mathbf{a}$ & $\Vert\mathbf{a}\Vert$: Euclidean norm; $\Vert\mathbf{a}\Vert_0$: zero norm; $\operatorname{iFFT}[\mathbf{a}]$: iFFT;  $\hat{\mathbf{a}}$: estimate of $\mathbf{a}$.\\
scalars $a$, $b$ & $a^*$: complex conjugate; $\operatorname{acos}[a]$: arcus cosine; {$\lfloor a \rceil$: rounding to nearest integer}; $\operatorname{min}[a, b]$: minimum of $a$ and $b$; $\operatorname{mod}[a, b]$: $a$ modulo $b$.\\
constants & $e$: Euler's number; $j$: imaginary unit.\\
			\bottomrule
		\end{tabular}}
	\end{center}
   \label{tab:notation}
\end{table} 

\subsection{Signal and Propagation Model}
\label{sec:sigpropmodel}

Let the microphone signals be critically sampled with period $T$, 
let $\omega$ denote the radial frequency and $M$ the number of microphones, and let $y_m(\omega)$ with $m \in \{1,\dots, M\}$ denote the $m^{\text{th}}$  microphone signal in the frequency domain.
Critical sampling implies that the microphone signals are bandlimited by 
\begin{align}
\omega_{\mathrm{b}} = \pi/T, \label{eq:w_0}
\end{align}
{that is $y_m(\omega) = 0$ for $\omega \geq \omega_{\mathrm{b}}$.}

In SRP, the DSB is usually not directly applied to the microphone signals, but to a frequency-weighted version thereof in order to improve spatial resolution \cite{dibiase2000high, dibiase2001robust, knapp1976generalized, velasco2016proposal, zhang2008does}. 
To this end, 
{the deterministic frequency-domain GCC (FD GCC) matrix $\bPsi(\omega) \in \mathbb{C}^{M\times M}$ is defined as}
\begin{align}
[\bPsi(\omega)]_{m,m'} &= \psi_{m,m'}(\omega)\nonumber\\
&=\gamma_{m,m'}(\omega)y_m(\omega)y_{m'}^*(\omega), \label{corrmatrix}
\end{align}
where $ \gamma_{m,m'}(\omega)$ denotes the weighting function.
A common, heuristically motivated weighting approach is the so-called phase transform (PHAT) \cite{knapp1976generalized, dibiase2000high, dibiase2001robust, zotkin2004accelerated, Silverman05, dmochowski2007generalized, do07, do2007fast, tervo2008interpolation, do09, cobos2010modified, marti13, nunes14, lima2014volumetric, velasco2016proposal, salvati2017exploiting, Coteli2018, grondin2019svd, DiazGuerra2021, GarciaBarrios2021, dietzen2021low, Cakmak2022, Grinstein2023, zhang2008does} given by $\gamma_{m,m'}(\omega) = {\displaystyle 1}/{\displaystyle \lvert y_m(\omega)y_{m'}^*(\omega)\rvert}$, in which case only phase information remains on the right-hand side of (\ref{corrmatrix}). 
The remainder of this paper holds regardless of the actual weighting approach.

 We consider both near- and far-field  (NF and FF) sound propagation.
Let the Cartesian coordinate vectors of the $M$ microphone locations be denoted as ${\mathbf{r}}_{m} \in \mathbb{R}^3$.
In case of NF propagation, let $\mathbf{q}_{i}  \in \mathbb{R}^3$ with $i \in \{1, \dots, J\}$ be the Cartesian coordinate vectors of the $J$ candidate locations. 
In case of FF propagation, let $\mathbf{q}_{i}$ instead denote the unit-length incident direction vectors of the $J$ candidate locations.
The TDOA $\Delta t_{m,m'}(i)$ of a sound wave originating from $\mathbf{q}_{i}$ as observed at the microphone pair $(m,m')$ is then given by
\begin{align}
 \Delta &t_{m,m'}(i) = \nonumber\\
 &\frac{1}{c}
 \begin{cases}
 \lVert {\mathbf{r}}_{m} - \mathbf{q}_{i} \rVert - \lVert  {\mathbf{r}}_{m'} - \mathbf{q}_{i} \rVert  & \text{for NF propagation,}\\
 ({\mathbf{r}}_{m} -  {\mathbf{r}}_{m'})^\transp\mathbf{q}_{i}  & \text{for FF propagation,}
\end{cases}
 \label{eq:Deltat_i}
\end{align}
where $c$ is the speed of sound.
Note that with  $d_{m,m'} = \lVert  {\mathbf{r}}_{m} - {\mathbf{r}}_{m'} \rVert$ the inter-microphone distance, we find that 
\begin{align}
\Delta t_{m,m'}(i)   &\in [-\Delta t_{m,m'|0}  ,\, \Delta t_{m,m'|0} ] \label{eq:Deltat_i_range} \\
\text{with}\quad\Delta t_{m,m'|0} &=d_{m,m'} /c,
\end{align}
where the limits of the interval are obtained if ${\mathbf{r}}_{m}$, ${\mathbf{r}}_{m'}$, and  $\mathbf{q}_i$ lie on the same line in the NF case or if  ${\mathbf{r}}_{m} - {\mathbf{r}}_{m'}$ and $\mathbf{q}_{i}$ are colinear in the FF case. 
The number of different microphone pairs $(m,m')$ is given by
\begin{align}
P = M(M-1)/2.
\end{align}
For convenience, we introduce the microphone pair index 
$p \in \{1, \dots, P\}$ and admit the alternative notation $\psi_p(\omega) = \psi_{m,m'}(\omega)$ and $\Delta t_p(i) = \Delta t_{m,m'}(i)$ for $m>m'$.

\subsection{Steered Response Power}
\label{sec:bfperspective}

The DSB steering vector $\h(\omega, i) \in \mathbb{C}^{M}$  \cite{brandstein2013microphone, gannot2017consolidated, SourceSep2018} 
towards the $i^\text{th}$ candidate location relative to the first microphone is solely 
defined by  $\Delta t_{m,1}(i)$ as
\begin{align}
[\h(\omega, i)]_m &= e^{-j\omega\Delta  t_{m,1}(i)}. \label{h_steering}
\end{align}
With $\bPsi(\omega)$ in (\ref{corrmatrix}) and $\h(\omega, i)$ in (\ref{h_steering}), we can define the
 frequency-dependent SRP map
\begin{align}
z(\omega,i) &= \h^\herm(\omega, i)\bPsi(\omega)\h(\omega, i) - \tr[\bPsi(\omega)]\nonumber\\
&= 2
\sum_{\scriptscriptstyle m'=1}^{\scriptscriptstyle M} 
\hspace{-.15cm}\sum_{\hspace{.2cm} \scriptscriptstyle m = m'+1 \hspace{-.2cm}}^{\scriptscriptstyle M}\hspace{-.15cm}
\Re\bigl[\psi_{m,m'}(\omega)e^{j\omega\Delta t_{m,m'}(i)}\bigr] \nonumber\\
&= 2
\sum_{p=1}^P 
\Re\bigl[\psi_{p}(\omega)e^{j\omega\Delta t_{p}(i)}\bigr], 
\label{SRP_omegaT}
\end{align}
where $\tr[\bPsi(\omega)]$ is commonly subtracted  \cite{dibiase2000high, dibiase2001robust, zotkin2004accelerated, Silverman05, dmochowski2007generalized, do07, do2007fast, tervo2008interpolation, do09, cobos2010modified, marti13, nunes14, lima2014volumetric, velasco2016proposal, salvati2017exploiting, Coteli2018, grondin2019svd, DiazGuerra2021, GarciaBarrios2021, dietzen2021low, Cakmak2022, Grinstein2023} as it constitutes an offset independent of $i$ and hence does not contribute to the objective of spatial mapping.
The second transition in (\ref{SRP_omegaT}) is obtained from the relation $\Delta t_{m,1}(i) - \Delta t_{m',1}(i) = \Delta t_{m,m'}(i)$ and the Hermitian structure of $\bPsi(\omega)$.
Here, the sum is to be taken over all combinations $(m,m')$ with $m>m'$, i.e. over all $P$ microphone pairs.
From (\ref{SRP_omegaT}), the broadband SRP value $z(i)$ is obtained by integrating $z(\omega, i)$ over frequency, 
\begin{align}
z(i) &= \int_{0}^{\omega_{\mathrm{b}}}\hspace{-.15cm}z(\omega, i)\,d\omega, \label{SRP_T}
\end{align}
where the upper integration limit coincides with the assumed bandlimit. 

The NF or FF location $\mathbf{q}_{\mathrm{s}}$ of a single source can be estimated from the candidate location index $i$ that maximizes $z(i)$, i.e. 
\begin{align}
\hat{\mathbf{q}}_{\mathrm{s}} &= \mathbf{q}_{i_{\mathrm{max}}} \label{eq:qshat}\\
\text{with} \quad i_{\mathrm{max}} &= \operatorname{arg}\underset{i}{\operatorname{max}}\, z(i). \label{eq:imax}
\end{align}

\subsection{Relation to TD GCC}
\label{sec:timedomainGCC}
The integral in (\ref{SRP_T}) may be reformulated into a simple relation between $z(i)$ and the TD GCCs of the $P$ microphone pairs \cite{dibiase2000high, dibiase2001robust, velasco2016proposal}.
Let $\xi_{p}(\tau)$ denote the TD GCC defined by the inverse Fourier transform of the FD GCC $\psi_{p}(\omega)$, i.e.
\begin{align}
\xi_{p}(\tau) &= \int_{-\omega_{\mathrm{b}}}^{\omega_{\mathrm{b}}}\hspace{-.15cm}\psi_{p}(\omega)e^{j\omega\tau}\,d\omega \nonumber \\
&= 2\int_{0}^{\omega_{\mathrm{b}}}\hspace{-.15cm}\Re[\psi_{p}(\omega)e^{j\omega\tau}]\,d\omega,
 \label{GCC_IFT}
\end{align}
where the second transition is valid because $\psi_{{p}}(\omega)$ is complex symmetric over $\omega$.
Then, using (\ref{SRP_omegaT}) in (\ref{SRP_T}) and comparing to (\ref{GCC_IFT}) one finds that $z(i)$ becomes
\begin{align}
z(i) &= \sum_{p=1}^P \xi_{p}(\Delta t_{p}(i)),
 \label{SRP_T_IFT}
\end{align}
i.e. a summation of TD GCCs $\xi_{p}(\tau)$ at lags $\tau$ corresponding to the candidate-location TDOAs $\Delta t_{p}(i)$.
Regarding (\ref{GCC_IFT}) and (\ref{SRP_T_IFT}), recall that $\Delta t_{p}(i) \in [-\Delta t_{p|0} ,\, \Delta t_{p|0}]$ according to (\ref{eq:Deltat_i_range}), while $\xi_{p}(\tau)$ itself has unlimited support in $\tau$.

\subsection{Matrix-Vector Form {of Conventional FD-based SRP}}
\label{sec:discreteFreq}

In the following, we represent the computation of the SRP map as a matrix-vector product using discrete frequencies. %
The formulation is based on the TD GCC formulation in Sec. \ref{sec:timedomainGCC}.
We assume that the discrete version of the FD GCC is computed from a discrete Fourier transform (DFT) of the original signals.
If the DFT is of length $2K$, we obtain discrete frequencies 
\begin{align}
\omega_k &= k \omega_{\mathrm{b}}/K \label{eq:w_k}
\end{align}
with the two-sided frequency bin index $k \in \{-K+1,\dots,K\}$. 
Noting that $\psi_{p}(\pm\omega_{\mathrm{b}}) = 0$ due to the bandlimit in (\ref{eq:w_0}),
the integral in (\ref{GCC_IFT}) can then be replaced\footnote{{Please note that although $\omega_k$ is discrete, $\tau$ in (\ref{eq:xi_p_discreteFreq}) is still continuous.}} by
\begin{align}
\xi_{p}(\tau) &=\sum_{k = -K+1}^{K-1}\psi_{p}(\omega_k)e^{j\omega_k\tau}\nonumber\\
&= 2\sum_{k = 1}^{K-1} \Re\bigl[ \psi_{p}(\omega_k)e^{j\omega_k\tau} \bigr],
\label{eq:xi_p_discreteFreq}
\end{align}
where the second transition is valid because $\psi_{{p}}(\omega)$ is complex symmetric and $\psi_{p}(0)$ is assumed to be negligible. 

Based on (\ref{SRP_T_IFT}) and the second, single-sided equality in (\ref{eq:xi_p_discreteFreq}), with the single-sided frequency bin index $k \in \{1,\dots,K-1\}$, we define the SRP map vector $\mathbf{z} \in \mathbb{R}^J$, the TD GCC vector 
$\boldsymbol{\xi}_{\Delta t_p} \in \mathbb{R}^J$, 
the single-sided
FD GCC vector $\bpsi_p \in \mathbb{C}^{K-1}$, and the single-sided SRP matrix $\mathbf{H}_p \in \mathbb{C}^{J \times K-1}$  as
\begin{align}
[\mathbf{z}]_i &= z(i),\label{eq:z_def}\\
\bigl[\boldsymbol{\xi}_{\Delta t_p} \bigr]_{i} &= \xi_{p}(\Delta t_{p}(i)),\label{eq:xi_Delta_p_single}\\
[\bpsi_p]_{k} &= \psi_{p}(\omega_k),\label{eq:psi_p_single}\\
[\mathbf{H}_p]_{i,k} &= e^{j\omega_k\Delta t_{p}(i)}.\label{eq:H_p_single}
\end{align}
Stacking over microphone pairs, we further define $\bpsi \in \mathbb{C}^{P(K-1)}$ and $\mathbf{H} \in \mathbb{C}^{J \times P(K-1)}$ as
\begin{align}
\bpsi &= \begin{pmatrix}
\bpsi^\transp_{1} & \cdots & \bpsi^\transp_{P} 
\end{pmatrix}^\transp, \label{eq:psi_single}\\
\mathbf{H} &= \begin{pmatrix}
\mathbf{H}_1 & \cdots & \mathbf{H}_P 
\end{pmatrix}. \label{eq:H_single}
\end{align}
Using the definitions in (\ref{eq:z_def})--(\ref{eq:H_single}), $\mathbf{z}$ can be expressed as
\begin{empheq}[box=\fbox]{align}
\mathbf{z} &= \sum_{p=1}^P \boldsymbol{\xi}_{\Delta t_p}\nonumber\\
&= 2 \sum_{p=1}^P \Re[\mathbf{H}_p \bpsi_p]\nonumber\\
&= 2\Re[\mathbf{H}  \bpsi]. \label{eq:z_computed}
\end{empheq} 
In the remainder, we refer to (\ref{eq:z_computed}) as conventional {FD-based} SRP (conv. FD-SRP)  and denote the corresponding source-location estimate by $\hat{\mathbf{q}}_{\mathrm{s}}$.
\subsection{{Discrete-time Approximation in Conventional TD-SRP}}
\label{sec:tdsrp}

{As an alternative to (\ref{eq:z_computed}), SRP is commonly computed using discrete samples \cite{dibiase2000high, dibiase2001robust, Silverman05, tervo2008interpolation} of $\xi_{p}(\tau)$ at $\tau = nT$ with $n \in \mathbb{Z}$, which can be obtained from the iFFT of $\psi_p(\omega_k)$.
Here, the term $\Delta t_{p}(i)$ in (\ref{SRP_T_IFT}) is often rounded to the nearest $nT$, i.e. (\ref{SRP_T_IFT}) is approximated by
\begin{empheq}[box=\fbox]{align}
z_{\operatorname{td}}(i) &= \sum_{p=1}^P \xi_{p}(      \lfloor \Delta t_{p}(i)/T\rceil  T  ).
 \label{eq:TD_SRP}
\end{empheq} 
In the remainder, we refer to (\ref{eq:TD_SRP}) as conventional TD-based SRP (conv. TD-SRP), define $\mathbf{z}_{\operatorname{td}}$ as $[\mathbf{z}_{\operatorname{td}}]_i = z_{\operatorname{td}}(i)$, and denote the corresponding source-location estimate by $\hat{\mathbf{q}}_{\mathrm{s}|\mathrm{td}}$.
While conv. TD-SRP is computationally significantly more efficient as compared to conv. FD-SRP as will be shown in Sec. \ref{sec:compcomp}, the approximation due to rounding can potentially reduce the localization performance in  scenarios with small error tolerances or small microphone array apertures, which is verified in the simulations in Sec. \ref{sec:locAccThreshAperture}.}

\subsection{Computational Complexity}
\label{sec:compcomp}
We define computational complexity as the number of real multiply-accumulate (MAC) operations {required during runtime to compute an SRP map given the FD GCC $\bpsi$.}\footnote{
{The complexity of computing $\bpsi$ using the FFT is in the order of $MK\log_2[2K]$  \cite{cormen2022introduction}, which can be verified to be at least two or three orders of magnitude below the complexity of the matrix-vector product in (\ref{eq:z_computed}) in usual setups, cf. also the parameters in Table \ref{tab:simparameters}.
As this step is common to all presented approaches, it is therefore neglected in the definition of complexity.}
} 

For a matrix-vector product, the number of MAC operations can be easily verified to be the number of (non-zero) elements of the matrix {\cite{cormen2022introduction, Meyer00}}.
The computation of the conv. {FD-}SRP map $\mathbf{z}$  in (\ref{eq:z_computed}) requires the real part of one complex matrix-vector product, namely $\mathbf{H} \cdot \bpsi$.
{Since $\mathbf{H} \in \mathbb{C}^{J \times P(K-1)}$ has non-zero elements only, the number of complex MAC operations is determined by the product of the matrix dimensions, i.e. $JP(K-1)$.
As the real part of one complex product is equivalent to two real products, we assume an additional factor $2$.}
The operation in (\ref{eq:z_computed}) therefore requires 
\begin{empheq}[box=\fbox]{align}
C = 2JP(K-1) \label{eq:SRPomp}
\end{empheq}
{MAC operations}.
Both $J$ and $K$ are design parameters and commonly chosen in the order of hundreds or thousands, while $P$ grows quadratically with $M$, making conv. {FD-}SRP rather complex. 
For an example, consider the parameters in Table \ref{tab:simparameters} chosen for the NF and FF scenario used in Sec. \ref{sec:sim}.

{The computation of the conv. {TD-}SRP map $\mathbf{z}_{\operatorname{td}}$  in (\ref{eq:TD_SRP}) requires first $P$ iFFTs of size $2K$ to obtain the samples $\xi_{p}(nT)$ from $\psi_p(\omega_k)$, and second the  accumulation of the $P$ samples $ \xi_{p}(      \lfloor \Delta t_{p}(i)/T\rceil  T  )$ for each of the $J$ candidate locations.
One iFFT of size $2K$ requires $2K\log_2[2K]$ complex {MAC operations}  \cite{cormen2022introduction}, where we assume an additional factor $4$.
The accumulation of all $ \xi_{p}(      \lfloor \Delta t_{p}(i)/T\rceil  T  )$ requires $JP$ MAC operations. 
The operation in (\ref{eq:TD_SRP}) therefore requires 
\begin{empheq}[box=\fbox]{align}
C_{\mathrm{td}} = JP + 8K\log_2[2K] \label{eq:tdSRPcomp}
\end{empheq}
MAC operations, which is less than (\ref{eq:SRPomp}) in usual setups, cf. the parameters in Table \ref{tab:simparameters}.
}

\section{Review of the State of the Art}
\label{sec:review}

We review two {low-complexity approximations of conv. FD-SRP} {yielding fine SRP maps}. 
A low-rank SRP approach referred to as SVD-PHAT \cite{grondin2019svd} is discussed in Sec. \ref{sec:LRSRP}, and the approach based on Nyquist-Shannon sampling and interpolation introduced in our prior work \cite{dietzen2021low} is discussed in Sec. \ref{sec:SISRP}.

\subsection{Low-Rank SRP}
\label{sec:LRSRP}

We discuss the low-rank approximation {of FD-SRP} in Sec. \ref{sec:LRSRPapprox} and the computational complexity of the approach in Sec. \ref{sec:LRSRPcc}.

\subsubsection{Low-Rank Approximation}
\label{sec:LRSRPapprox}

\newcommand{\tall}{\mathrm{tall}}
\newcommand{\fat}{\mathrm{fat}}

SVD-PHAT \cite{grondin2019svd} can be defined based on the cost function
\begin{align}
f(\mathbf{H}_{\mathrm{var}}) &= \Vert \mathbf{H}_{\mathrm{var}} - \mathbf{H}\Vert_{\mathrm{F}}^2 \label{eq:f_lr}
\end{align}
where $\mathbf{H}_{\mathrm{var}}$ is the optimization variable. 
In order to obtain a computationally less demanding approximation ${\mathbf{H}}_{\mathrm{lr}}$ of $\mathbf{H}$, (\ref{eq:f_lr}) is minimized subject to a low-rank constraint, i.e. 
\begin{align}
{\mathbf{H}}_{\mathrm{lr}} &= \arg\underset{\mathbf{H}_{\mathrm{var}}}{\min}\, \Vert \mathbf{H}_{\mathrm{var}} - \mathbf{H}\Vert_{\mathrm{F}}^2  \, \operatorname{s.t.}   \, \mathrm{rank}[\mathbf{H}_{\mathrm{var}}] \leq R_H, \label{eq:min_lr}
\end{align}
where the choice of the design parameter $R_H < \operatorname{min}[J, P(K-1)]$ determines the computational complexity, cf. Sec. \ref{sec:LRSRPcc}.
The solution to (\ref{eq:min_lr}) is based on the singular value decomposition  (SVD) of $\mathbf{H}$  \cite{Meyer00} and can be written as an outer product of a tall matrix ${\mathbf{H}}_{\tall} \in \mathbb{C}^{J \times R_H}$ and a fat matrix $ {\mathbf{H}}_{\fat} \in \mathbb{C}^{R_H \times P(K-1)}$, i.e. 
\begin{align}
{\mathbf{H}}_{\mathrm{lr}} &= {\mathbf{H}}_{\tall} {\mathbf{H}}_{\fat}, \label{eq:H_lr}
\end{align}
with for instance ${\mathbf{H}}_{\tall} = {\mathbf{U}}\tilde{\mathbf{\Sigma}}$ and ${\mathbf{H}}_{\fat} = {\mathbf{V}}^\herm$, where the diagonal of $\tilde{\mathbf{\Sigma}} \in \mathbb{R}^{R_H \times R_H}$ contains the $R_H$ largest singular values of $\mathbf{H}$, 
and the columns of ${\mathbf{U}} \in \mathbb{C}^{J \times R_H}$ and ${\mathbf{V}} \in \mathbb{C}^{P(K-1)\times R_H}$ are given by the corresponding left and right singular vectors, respectively.

The corresponding SRP map is defined similarly to (\ref{eq:z_computed}) as
\begin{empheq}[box=\fbox]{align}
\mathbf{z}_{\mathrm{lr}} &=2\Re[{\mathbf{H}}_{\mathrm{lr}} \bpsi] \nonumber\\
&=2\Re[{\mathbf{H}}_{\tall}  {\mathbf{H}}_{\fat} \bpsi] \label{eq: z_lr}.
\end{empheq}
In the remainder, we refer to (\ref{eq: z_lr}) as low-rank SRP (LR-SRP) and denote the corresponding source-location estimate by $\hat{\mathbf{q}}_{\mathrm{s}|\mathrm{lr}}$.

\subsubsection{Computational Complexity}
\label{sec:LRSRPcc}

The decomposition of ${\mathbf{H}}_{\mathrm{lr}}$ in (\ref{eq:H_lr}) can be computed upfront and hence does not require computations during run time.
The computation of the LR-SRP map $\mathbf{z}_{\mathrm{lr}}$ in (\ref{eq: z_lr}) requires two matrix-vector products, namely first ${\mathbf{H}}_{\fat} \cdot \bpsi$ resulting in the vector ${\mathbf{H}}_{\fat}\bpsi$, and second ${\mathbf{H}}_{\tall} \cdot  {\mathbf{H}}_{\fat}\bpsi$.
The computational complexity thereof is determined by the number of (non-zero) elements in ${\mathbf{H}}_{\fat}$ and ${\mathbf{H}}_{\tall}$, given by $R_HP(K-1)$ and $JR_H$, respectively.
Both products are complex, but of the second product only the real part is required, such that we assume additional factors of $4$ and $2$, respectively.
The operation in (\ref{eq: z_lr}) therefore requires 
\begin{empheq}[box=\fbox]{align}
C_{\mathrm{lr}} = 2JR_H + 4R_HP(K-1) \label{eq:C_lr}
\end{empheq}
{MAC operations}. 
If $R_H$ is chosen small enough, $C_{\mathrm{lr}}$ will be well below $C$ in (\ref{eq:SRPomp}).

\subsection{Sampling and Interpolation-based SRP}
\label{sec:SISRP}

We discuss sampling and interpolation according to \cite{dietzen2021low} in Sec. \ref{sec:samplinginterpolation}, a corresponding matrix-vector form in Sec. \ref{sec:discreteFreq2}, and the computational complexity of the approach in Sec. \ref{sec:SISRPcc}.

\subsubsection{Sampling and Interpolation}
\label{sec:samplinginterpolation}

Recall that $\xi_{p}(\tau)$ is bandlimited by $\omega_{\mathrm{b}}$, as can be seen from (\ref{GCC_IFT}).
We can therefore invoke the Nyquist-Shannon sampling theorem \cite{marks2012introduction}.  
Let $\xi_{p}(\tau)$ be critically sampled at $\tau = nT$ with $n \in \mathbb{Z}$.
Using Whittaker-Shannon interpolation  \cite{marks2012introduction}, we can then express $\xi_{p}(\tau)$ as a function of its samples $\xi_{p}(nT)$, 
\begin{align}
\xi_{p}(\tau) = \sum_{n = -\infty}^{\infty} \xi_{p}(nT) \sinc(\tau/T - n), \label{eq:SRPinterp}
\end{align}
permitting perfect reconstruction. 
As $\xi_{p}(\tau)$ is bandlimited in $\omega$, it has unlimited support in $\tau$  \cite{marks2012introduction} and thus, as stated by (\ref{eq:SRPinterp}), one indeed needs infinitely many samples  $\xi_{p}(nT)$ in order to exactly reconstruct $\xi_{p}(\tau)$.
Nonetheless, we may approximate $\xi_{p}(\tau)$ based on (\ref{eq:SRPinterp}) and the following arguments.

In the SRP use case, the approximation needs to be accurate only within the interval $\tau \in[-\Delta t_{p|0} ,\, \Delta t_{p|0}]$,  cf. (\ref{eq:Deltat_i})--(\ref{eq:Deltat_i_range}), where $\tau$ may be interpreted as the TDOA from a physical source location. 
Samples $\xi_{p}(nT)$ sufficiently far {outside} this interval, i.e. at $\lvert n \rvert  \gg \Delta t_{p|0}/T$, have little effect {within} the interval for two reasons. 
First, their effect is limited by nature of the interpolating sinc-function, whose envelope decays with the inverse of its argument.
Second, also the samples $\xi_{p}(nT)$ themselves decay 
as the TD GCC lag $nT$ does not correspond to a TDOA anymore.\footnote{Exceptions due to, e.g., strong reflections or source-signal periodicities are possible, but irrelevant to the source localization problem.} 
In \cite{dietzen2021low}, we have therefore proposed to approximate $\xi_{p}(\tau)$ by $\xi_{p|\mathrm{si}}(\tau)$ by truncating (\ref{eq:SRPinterp}) as
\begin{align}
\xi_{p|\mathrm{si}}(\tau) &=\sum_{n = -N_{p}}^{N_{p}} \xi_{p}(nT) \text{sinc}(\tau/T - n),\label{eq:SRPapprox}\\
\text{with}\,\,\,\,\, N_{p} & = \left\lfloor \Delta t_{p|0}/T \right\rfloor + N_{\mathrm{aux}}, \label{eq:N_p}
\end{align}
where 
$ N_{\mathrm{aux}}$ is a design parameter defining the number of auxiliary sample points beyond $[-\Delta t_{p|0} ,\, \Delta t_{p|0}]$. 
In practice, few auxiliary samples are sufficient, cf. also Sec.~\ref{sec:sim} and \cite{dietzen2021low}.
Based on the above, the Nyquist-Shannon approximation of the SRP map is obtained by first computing the required samples $\xi_{p}(nT)$ based on (\ref{eq:xi_p_discreteFreq}), then interpolating $\xi_{p|\mathrm{si}}(\tau)$ at $\tau = \Delta t_p(i)$ according to (\ref{eq:SRPapprox}), and finally accumulating over $p$ according to (\ref{SRP_T_IFT}). 
This procedure will become more apparent in the following subsection \ref{sec:discreteFreq2}.

\subsubsection{Matrix-Vector Form}
\label{sec:discreteFreq2}

In the following, we represent the computation of the sampling and interpolation-based SRP map as a matrix-vector product.
This respresentation has not been included in our previous work \cite{dietzen2021low}, but is essential to the approaches proposed in Sec. \ref{sec:lr_sp_int}.\footnote{Despite its novel aspect, we introduce the  matrix-vector form in the state-of-the-art section to ensure a coherent outline and facilitate the complexity analysis in the following section.}

With $n \in \{-N_p,\dots,N_p\}$, let the column or row index ${n}_{\mathrm{mod}}$ be defined as
\begin{align}
{n}_{\mathrm{mod}} &= \operatorname{mod}[n, 2N_p+1]+1,
\label{eq:nmod}
\end{align}
yielding ${n}_{\mathrm{mod}} \in \{1,\dots,2N_{p}+1\}$.
Further, let $N$ denote the average number of required samples per microphone pair, i.e.
\begin{align}
N= 1 + \frac{2}{P}\sum_{p=1}^{P}{N_p}. \label{eq:N}
\end{align}
Based on (\ref{eq:xi_p_discreteFreq}), we define the TD GCC sample vector $\boldsymbol{\xi}_{p} \in \mathbb{R}^{2N_p + 1}$ and the single-sided sampling matrix $\mathbf{S}_p \in \mathbb{C}^{2N_p + 1 \times K-1}$ by 
\begin{align}
[\boldsymbol{\xi}_{p}]_{{n}_{\mathrm{mod}}} &= \xi_p(nT), \label{eq:eq:xi_p_single}\\
[\mathbf{S}_p]_{{n}_{\mathrm{mod}},k} &= e^{j  \pi k n/K}, \label{eq:S_p_single}
\end{align}
which can be compared to (\ref{eq:xi_Delta_p_single}) and (\ref{eq:H_p_single}).
In (\ref{eq:S_p_single}), we have used $\omega_k n T =  \pi k n/K$ from (\ref{eq:w_0}) and (\ref{eq:w_k}).
Note that $\mathbf{S}_p$ is a submatrix of the $2K$ inverse DFT (iDFT) matrix corresponding to a subset of samples and one side of the spectrum.
Stacking over microphone pairs, we define $\boldsymbol{\xi}  \in \mathbb{R}^{PN}$ and $\mathbf{S} \in \mathbb{C}^{PN \times P(K-1)}$ as
\begin{align}
\boldsymbol{\xi} &= \begin{pmatrix}
\boldsymbol{\xi}_{1}^\transp &\cdots &\boldsymbol{\xi}_{P}^\transp
\end{pmatrix}^\transp \label{eq:xi_single}\\
\mathbf{S} &= 
\operatorname{blkdiag} \bigl[
\mathbf{S}_1,\, \dots,\, \mathbf{S}_P
\bigr],\label{eq:S_single}
\end{align}
Using (\ref{eq:eq:xi_p_single}) to (\ref{eq:S_single}), with $\bpsi$ defined in (\ref{eq:psi_single}), we find that $\boldsymbol{\xi}_p = 2\Re[\mathbf{S}_p  \bpsi_p]$ and accordingly 
\begin{align}
\boldsymbol{\xi} &= 2\Re[\mathbf{S}  \bpsi]. \label{eq:xi_computed}
\end{align} 
Based on (\ref{eq:SRPapprox}), we define the interpolated TD GCC vector $\boldsymbol{\xi}_{\Delta t_p|\mathrm{si}}  \in \mathbb{R}^{J}$ and the interpolation matrix $\bLambda_p  \in \mathbb{R}^{J \times (2N_p + 1)}$ as
\begin{align}
\bigl[\boldsymbol{\xi}_{\Delta t_p|\mathrm{si}} \bigr]_{i}  &=\xi_{p|\mathrm{si}}(\Delta t_{p}(i)),
\label{eq:xi_Delta_p_single_interp}\\
[\bLambda_p]_{i,{n}_{\mathrm{mod}}} &= \operatorname{sinc}(\Delta t_{p}(i)/T - n). \label{eq:Lambda_p}
\end{align}
Stacking over microphone pairs, we define $\bLambda \in \mathbb{R}^{J \times PN}$ as
\begin{align}
\bLambda &= \begin{pmatrix}
\bLambda_1 &\cdots & \bLambda_P
\end{pmatrix}. \label{eq:Lambda}
\end{align}
Using (\ref{eq:xi_Delta_p_single_interp}) to (\ref{eq:Lambda}), with $\boldsymbol{\xi}_{p}$ and  $\boldsymbol{\xi}$ defined in (\ref{eq:eq:xi_p_single}) and  (\ref{eq:xi_single}), we find that
 $\boldsymbol{\xi}_{\Delta t_p|\mathrm{si}} = \bLambda_p\boldsymbol{\xi}_{p}$
 and accordingly define the corresponding SRP map ${\mathbf{z}_\mathrm{si}}$ based on (\ref{SRP_T_IFT}) as
\begin{empheq}[box=\fbox]{align}
{\mathbf{z}_\mathrm{si}} &= \sum_{p=1}^P\boldsymbol{\xi}_{\Delta t_p|\mathrm{si}} \nonumber \\
&= \sum_{p=1}^P \bLambda_p\boldsymbol{\xi}_{p} \nonumber  \\
&=\bLambda \boldsymbol{\xi}. \label{eq:z_circ_computed}
\end{empheq}
With (\ref{eq:xi_computed}) and (\ref{eq:z_circ_computed}), we have approximated the {conv. FD-SRP matrix} $\mathbf{H}$ in (\ref{eq:z_computed}) by
\begin{align}
\mathbf{H}_{\mathrm{si}}  &= \bLambda\mathbf{S}.
\label{eq:H_circ}
\end{align}
Note that like (\ref{eq:H_lr}), we can under most cases of practical interest interpret (\ref{eq:H_circ}) as low-rank: with  $\bLambda \in \mathbb{R}^{J \times PN}$ and $\mathbf{S} \in \mathbb{C}^{PN \times P(K-1)}$ we typically find that $\bLambda$ is tall and 	$\mathbf{S}$ is fat since in usual setups $J > NP$ and $N < K-1$,\footnote{In order to ensure a good TD GCC estimate within the TDOA range and limit the effects of windowing and circular convolution, the frame length $2K$ should well exceed twice the largest possible TDOA $ \Delta t_{p|0}$, which implies that $N \ll 2K$, cf. (\ref{eq:N}) and (\ref{eq:N_p}).} respectively, cf. also the parameters in Table \ref{tab:simparameters}.
{The approach in (\ref{eq:H_lr}) can also be compared to conv. TD-SRP in (\ref{eq:TD_SRP}), which is obtained if (\ref{eq:Lambda_p}) is replaced by $[\bLambda_p]_{i,{n}_{\mathrm{mod}}} = 1$ if  $\lfloor \Delta t_{p}(i)/T\rceil = n$ and $[\bLambda_p]_{i,{n}_{\mathrm{mod}}} = 0$ otherwise.}

In the remainder, we refer to (\ref{eq:z_circ_computed}) as sampling + interpolation-based SRP (SI-SRP) and denote the corresponding source-location estimate by $\hat{\mathbf{q}}_{\mathrm{s}|\mathrm{si}}$.

\subsubsection{Computational Complexity}
\label{sec:SISRPcc}

The components of $\mathbf{H}_{\mathrm{si}}$ in (\ref{eq:H_circ}) can be defined upfront and hence do not require computations during run time.
The computation of the SI-SRP map $\mathbf{z}_{\mathrm{si}}$  in (\ref{eq:z_circ_computed}) requires sampling and interpolation.
The sampling operation, i.e. the computation of  $\boldsymbol{\xi}$, can be performed in two different ways.
We may either use the matrix vector product $\mathbf{S} \cdot \bpsi$ in (\ref{eq:xi_computed}), whereof the complexity is determined by the number of non-zero elements
in the block-diagonal matrix $\mathbf{S}$, i.e.  $PN(K-1)$.
Since the product is complex, but only its real part is required, we assume an additional factor $2$. 
Alternatively, noting that $\mathbf{S}_p$ is a submatrix of the $2K$-iDFT matrix, we may use the iFFT {as in conv. TD-SRP}.\footnote{The use of the iFFT has not been considered in our previous work \cite{dietzen2021low}.}
Let $\bar{\bpsi}_p$ be a two-sided version of ${\bpsi}_p$ as defined in the Appendix.
Then, $\boldsymbol{\xi}_{p}$ can be obtained by performing the iFFT on $\bar{\bpsi}_p$ and selecting the required $2N_p + 1$ samples according to
\begin{align}
\bar{\boldsymbol{\xi}}_{p} &= \operatorname{iFFT}[\bar{\bpsi}_p],\label{eq:xi_p_ifft}\\
[\boldsymbol{\xi}_{p}]_{{n}_{\mathrm{mod}}} &= [\bar{\boldsymbol{\xi}}_{p}]_{{n}^{\prime}_{\mathrm{mod}}}, \label{eq:xi_p_from_ifft}
\end{align}
with ${n}_{\mathrm{mod}}$ in (\ref{eq:nmod}) and the sample selection index ${n}^{\prime}_{\mathrm{mod}}$ defined by
\begin{align}
{n}^{\prime}_{\mathrm{mod}} &= \operatorname{mod}[n, 2K]+1, \label{eq:n_prime} 
\end{align}
yielding ${n}^{\prime}_{\mathrm{mod}} \in \{1,\dots,N_{p}+1\} \cup \{2K-N_{p}+1,\dots,2K\}$ for $n \in \{-N_p,\dots,N_p\}$.
From $\boldsymbol{\xi}_{p}$ in (\ref{eq:xi_p_from_ifft}), we obtain the stacked TD GCC vector $\boldsymbol{\xi}$ using (\ref{eq:xi_single}).
One iFFT of size $2K$ requires $2K\log_2[2K]$ complex {MAC operations}, where we assume an additional factor $4$.
Computing $\boldsymbol{\xi}$ therefore requires $C_{\mathrm{samp}}$ {MAC operations} with 
\begin{align}
C&_{\mathrm{samp}} = \nonumber\\
&\begin{cases}
2PN(K-1) 				& \text{when using (\ref{eq:xi_computed})},\\
8PK\log_2[2K] 	& \text{when using (\ref{eq:xi_p_ifft}), (\ref{eq:xi_p_from_ifft})}, (\ref{eq:xi_single}). 
\end{cases}\label{eq:C_sampling}
\end{align}
The iFFT-based computation of $\boldsymbol{\xi}$ is less complex if $N > 4\frac{K}{K-1}\log_2[2K]$, which is often the case in usual setups, cf. the parameters in Table \ref{tab:simparameters}.

The interpolation operation requires one real matrix-vector product, namely $\bLambda \cdot \boldsymbol{\xi}$ in (\ref{eq:z_circ_computed}).
The computational complexity thereof is determined by the number of (non-zero) elements in $\bLambda$, given by $JPN$.
In total, the operation in (\ref{eq:z_circ_computed}) therefore requires 
\begin{empheq}[box=\fbox]{align}
C_{\mathrm{si}} = JPN + C_{\mathrm{samp}}
\end{empheq}
{MAC operations}.
For usual setups, we typically find that $C_{\mathrm{si}}$ will be well below $C$ of {conv. FD-SRP} in (\ref{eq:SRPomp}), cf. also the low-rank interpretation of (\ref{eq:H_circ}) and Table \ref{tab:simparameters}.
{In comparison to $C_{\operatorname{td}}$ of conv. TD-SRP in (\ref{eq:tdSRPcomp}), however, we observe that the interpolation operation of SI-SRP is $N$ times more complex than the corresponding sample selection of conv. TD-SRP.}  

\section{Low-Rank and Sparse Interpolation}
\label{sec:lr_sp_int}

The SI-SRP approach presented in Sec. \ref{sec:SISRP} reduces computational complexity depending on the acoustic scenario and algorithmic parameter choices, but is hardly scalable in the sense of trading-off complexity against performance. 
In this section, we therefore propose two scalable extensions of the SI-SRP approach.
While the sampling operation of the SI-SRP approach can already be carried out in a computationally efficient manner using the iFFT, {the complexity of the interpolation operation is $N$ times larger than the corresponding sample selection of conv. TD-SRP.}
We therefore further reduce complexity based on low-rank and sparse approximations of the interpolation matrix.
In Sec. \ref{sec:interpolApprox}, we derive the proposed approaches, and in Sec. \ref{sec:SSPLR_SRPcc} discuss the complexity of the resulting algorithms.

\subsection{Low-Rank and Sparse Interpolation-based SRP}
\label{sec:interpolApprox}

We define the cost function
\begin{align}
f(\bLambda_{\mathrm{var}}) &= \Vert \bLambda_{\mathrm{var}}\mathbf{S} - \mathbf{H}\Vert_{\mathrm{F}}^2
\label{eq:cost_orig}
\end{align}
where $\bLambda_{\mathrm{var}} \in \mathbb{R}^{J \times PN}$ is the optimization variable.
Similarly to (\ref{eq:f_lr}), the cost function in (\ref{eq:cost_orig}) penalizes the approximation error with respect to $ \mathbf{H}$.

To reduce complexity, (\ref{eq:cost_orig}) could be minimized subject to low-rank or sparsity constraints on $\bLambda_{\mathrm{var}}$.\footnote{In addition, one would either require a realness constraint on $\bLambda_{\mathrm{var}}$ or alternatively use a two-sided version of (\ref{eq:cost_orig}), cf. also the corresponding two-sided definitions $\bar{\mathbf{H}}$ and $\bar{\mathbf{S}}$ in the Appendix.} 
Such a problem could for instance be solved iteratively by means of the proximal gradient method \cite{antonello2020proximal}, that is by alternatingly performing a gradient-descent step based on the convex cost function in (\ref{eq:cost_orig}) and applying a proximal operator based on the constraints.
Instead however, we propose to simplify the cost function in order to obtain a simpler minimization problem that can be solved without requiring alternations.

To this end, we first approximate (\ref{eq:cost_orig}) by replacing $\mathbf{H}$ by $\mathbf{H}_{\mathrm{si}}$ yielding 
\begin{align}
f(\bLambda_{\mathrm{var}}) &\approx \Vert \bLambda_{\mathrm{var}}\mathbf{S} - \mathbf{H}_{\mathrm{si}}\Vert_{\mathrm{F}}^2,\label{eq:cost_fun_interp}
\end{align}
and then exploit the properties of $\mathbf{S}$ as outlined in the following.
With $\mathbf{H}_{\mathrm{si}} = \bLambda\mathbf{S}$ from (\ref{eq:H_circ}), we find that the right-hand side of (\ref{eq:cost_fun_interp}) becomes
\begin{align}
\Vert \bLambda_{\mathrm{var}}\mathbf{S} - \mathbf{H}_{\mathrm{si}}\Vert_{\mathrm{F}}^2 &= \Vert (\bLambda_{\mathrm{var}} - \bLambda)\mathbf{S}\Vert_{\mathrm{F}}^2\nonumber\\
&\approx \frac{1}{2} \Vert (\bLambda_{\mathrm{var}} - \bLambda)\bar{\mathbf{S}}\Vert_{\mathrm{F}}^2\nonumber\\
&= K\Vert \bLambda_{\mathrm{var}} - \bLambda \Vert_{\mathrm{F}}^2,
\label{eq:cost_simplified}
\end{align}
where $\bar{\mathbf{S}}$ is a two-sided version of $\mathbf{S}$ as defined in the Appendix and the second transition holds for $\bLambda_{\mathrm{var}} - \bLambda$ being real.
The error in this transition is solely due to the columns of $\bar{\mathbf{S}}$ relating to $\omega_k = 0$ and $\omega_k = \pm \omega_{\mathrm{b}}$, which are disregarded in the definition of $\mathbf{S}$, cf. (\ref{eq:S_p_single}) and (\ref{eq:S_single}), and is considered negligible.
The third transition holds because $\Vert \mathbf{A} \Vert_{\mathrm{F}}^2 = \operatorname{tr}[\mathbf{A}\mathbf{A}^\herm]$ for any matrix $\mathbf{A}$ and the rows of $\bar{\mathbf{S}}$ are orthogonal with $\bar{\mathbf{S}}\bar{\mathbf{S}}^\herm= 2K \mathbf{I}$.

Hence, the last equality in (\ref{eq:cost_simplified}) is nearly equivalent to (\ref{eq:cost_orig}) and can be used as a cost function instead.
In order to obtain computationally less demanding approximations $\bLambda_{\mathrm{lr}}$ and $\bLambda_{{\mathrm{sp}}}$ of $\bLambda$, we propose to minimize this cost function subject to low-rank or sparsity constraints, respectively, i.e.
\begin{align}
\bLambda_{\mathrm{lr}} &= \arg\underset{\bLambda_{\mathrm{var}}}{\min}\, \Vert \bLambda_{\mathrm{var}} - \bLambda \Vert_{\mathrm{F}}^2 \, \operatorname{s.t.} \, \mathrm{rank}[\bLambda_{\mathrm{var}}] \leq R_{\Lambda},\label{eq:min_Lambda_lr}\\
\bLambda_{{\mathrm{sp}}} &= \arg\underset{\bLambda_{\mathrm{var}}}{\min}\, \Vert \bLambda_{\mathrm{var}} - \bLambda \Vert_{\mathrm{F}}^2  \,  \operatorname{s.t.}   \,  
\Vert\mathrm{vec}[\bLambda_{\mathrm{var}}]\Vert_0 \leq Q_{\Lambda}, \label{eq:min_Lambda_sp}
\end{align}
where the choices of the design parameters $R_{\Lambda} < \operatorname{min}[J, PN]$ and $Q_{\Lambda} < JPN$ determine the computational complexity, cf. Sec. \ref{sec:SSPLR_SRPcc}.

The solution to (\ref{eq:min_Lambda_lr}) is based on the SVD of $\bLambda$  \cite{Meyer00} and can be written as an outer product of a tall matrix $\bLambda_{\tall} \in \mathbb{R}^{J \times R_{\Lambda}}$ and a fat matrix $ \bLambda_{\fat} \in \mathbb{R}^{R_{\Lambda} \times PN}$, i.e. 
\begin{align}
\bLambda_{\mathrm{lr}} = \bLambda_{\tall} \bLambda_{\fat}, \label{eq:Lambda_lr}
\end{align}
where $ \bLambda_{\tall}$ and $\bLambda_{\fat}$ can be defined similarly to ${\mathbf{H}}_{\tall}$ and ${\mathbf{H}}_{\fat}$ in Sec. \ref{sec:LRSRPapprox}.
The solution to (\ref{eq:min_Lambda_sp}) is obtained from $\bLambda$ by keeping the $Q_{\Lambda}$ largest elements and discarding all others. 

Using (\ref{eq:min_Lambda_lr}) to (\ref{eq:Lambda_lr}), with  $\boldsymbol{\xi}$ defined in (\ref{eq:xi_single}), we define the corresponding SRP maps ${\mathbf{z}_{\mathrm{slri}}}$ and $\mathbf{z}_{\mathrm{sspi}}$ similarly to (\ref{eq:z_circ_computed}) as
\begin{empheq}[box=\fbox]{align}
\mathbf{z}_{\mathrm{slri}} &= \bLambda_{\mathrm{lr}} \boldsymbol{\xi}\nonumber\\
&=\bLambda_{\tall} \bLambda_{\fat} \boldsymbol{\xi},\label{eq:z_circlr_computed}\\
\mathbf{z}_{\mathrm{sspi}} &= \bLambda_{\mathrm{sp}} \boldsymbol{\xi}. \label{eq:z_circsp_computed}
\end{empheq}
With (\ref{eq:xi_computed}) and  (\ref{eq:z_circlr_computed}) or (\ref{eq:z_circsp_computed}), we have approximated $\mathbf{H}$ in (\ref{eq:z_computed}) by
\begin{align}
\mathbf{H}_{\mathrm{slri}}  &= \bLambda_{\mathrm{lr}}\mathbf{S}, \label{eq:H_slri}\\ 
\mathbf{H}_{\mathrm{sspi}}  &= \bLambda_{\mathrm{sp}}\mathbf{S},  \label{eq:H_sspi}
\end{align}
respectively.
In the remainder, we refer to (\ref{eq:z_circlr_computed}) and (\ref{eq:z_circsp_computed}) as sampling + low-rank interpolation-based SRP (SLRI-SRP) and sampling + sparse interpolation-based SRP (SSPI-SRP) and denote the corresponding source-location estimate by $\hat{\mathbf{q}}_{\mathrm{s}|\mathrm{slri}}$ and $\hat{\mathbf{q}}_{\mathrm{s}|\mathrm{sspi}}$, respectively.

\subsection{Computational Complexity}
\label{sec:SSPLR_SRPcc}

{Because $\mathbf{H}$ in (\ref{eq:H_p_single})--(\ref{eq:H_single}) depends only the geometries of the microphone array and the search grid and the frequency resolution, the decompositions of $\bLambda_{\mathrm{lr}}$, $\mathbf{H}_{\mathrm{slri}}$, and $\mathbf{H}_{\mathrm{sspi}}$ in (\ref{eq:Lambda_lr}), (\ref{eq:H_slri}), and (\ref{eq:H_sspi}) can be computed upfront and hence do not require computations during run time.
The complexity of solving the optimization problems in (\ref{eq:min_Lambda_lr})--(\ref{eq:min_Lambda_sp}) is therefore not considered in our complexity analysis. }

The computation of the SLRI-SRP map $\mathbf{z}_{\mathrm{slri}}$ and the SSPI-SRP $\mathbf{z}_{\mathrm{sspi}}$  in (\ref{eq:z_circlr_computed}) and (\ref{eq:z_circsp_computed}) requires sampling and interpolation, where the complexity of the sampling operation $C_{\mathrm{samp}}$ remains unaltered and is given by (\ref{eq:C_sampling}).
The interpolation operation in (\ref{eq:z_circlr_computed}) requires two real matrix-vector products, namely first $\bLambda_{\fat} \cdot \boldsymbol{\xi}$ resulting in the vector $\bLambda_{\fat} \boldsymbol{\xi}$, and second $\bLambda_{\tall} \cdot \bLambda_{\fat} \boldsymbol{\xi}$. 
The computational complexity thereof is determined by the number of (non-zero) elements in $\bLambda_{\fat}$ and $\bLambda_{\tall}$, given by $R_{\Lambda}PN$ and $JR_{\Lambda}$, respectively.
The interpolation operation in (\ref{eq:z_circsp_computed}) requires one real matrix-vector product, namely $\bLambda_{\mathrm{sspi}} \cdot \boldsymbol{\xi}$, whereof the computational complexity is determined by the number of (non-zero) elements in $\bLambda_{\mathrm{sp}}$, i.e. $Q_{\Lambda}$.
In total, the operations in (\ref{eq:z_circlr_computed}) or (\ref{eq:z_circsp_computed}) therefore require 
\begin{empheq}[box=\fbox]{align}
C_{\mathrm{slri}} &= 
JR_{\Lambda}  + R_{\Lambda}PN + C_{\mathrm{samp}}\\
C_{\mathrm{sspi}} &=
Q_{\Lambda} + C_{\mathrm{samp}}
\end{empheq}
{MAC operations}.
If $R_{\Lambda}$ or $Q_{\Lambda}$ are chosen small enough, $C_{\mathrm{slri}}$ and $C_{\mathrm{sspi}}$ will be well below $C$ in (\ref{eq:SRPomp}), respectively.

\section{Simulations}
\label{sec:sim}

{We discuss the performance measures in Sec. \ref{sec:perm} and perform simulations using simulated room impulse responses (RIRs) in Sec. \ref{sec:simRIR} and measured RIRs in Sec. \ref{sec:measRIR}}. 

\definecolor{mycolor1}{rgb}{0.90000,0.25000,0.25000}%
\definecolor{mycolor2}{rgb}{0.25000,0.25000,0.90000}%
\definecolor{mycolor3}{rgb}{0.46600,0.67400,0.18800}%
\definecolor{mycolor4}{rgb}{0.00000,0.44700,0.74100}%
\definecolor{mycolor5}{rgb}{0.85000,0.32500,0.09800}%
\definecolor{mycolor6}{rgb}{0.92900,0.69400,0.12500}%
\definecolor{mycolor7}{rgb}{0.49400,0.18400,0.55600}%

\setlength\fwidth{7.12cm}
\setlength\fheight{.75\fwidth} 
 \begin{figure}
\centering
\hspace*{-0.19cm} 
    \input{./fig/geometry_nf.tex} 
\caption{Microphone [\ref{geometry_nf_mic}] and candidate locations [\ref{geometry_nf_cand}] in the NF scenario. The three-dimensional search grid samples a volume of $(4 \times 5 \times 0.1) \SI{}{m}$. For better visibility, the search grid is shown with a reduced resolution of \SI{10}{cm} instead of the simulated resolution of \SI{3.33}{cm} per spatial dimension. The axes limits correspond to the room dimensions.}
\label{fig:nf_geom}
\end{figure}

 \begin{figure}
\centering
\hspace*{-0.19cm} 
    \input{./fig/geometry_ff.tex} 
\caption{Microphone [\ref{geometry_nf_mic}] and candidate locations [\ref{geometry_nf_cand}] in the FF scenario with array diameter $\diameter=\SI{30}{cm}$. 
The two-dimensional search grid samples the lower half-sphere of incident directions, indicated at \SI{2}{m} distance from the center of the microphone array beyond which the FF assumption is assumed to be valid. 
For better visibility, the search grid is shown with a reduced resolution of 4$^\circ$ instead of the simulated resolution of 2$^\circ$ per angular dimension. The axes limits correspond to the room dimensions.}
\label{fig:ff_geom}
\end{figure}

\setlength\fwidth{16.74cm}
\setlength\fheight{4.75cm} 
 \begin{figure*}[t]
\centering
\hspace*{-0.18cm} 
%
%
\definecolor{mycolor1}{rgb}{0.00000,0.44700,0.74100}%
\definecolor{mycolor2}{rgb}{0.85000,0.32500,0.09800}%
\definecolor{mycolor3}{rgb}{0.92900,0.69400,0.12500}%
\begin{tikzpicture}

\begin{axis}[%
width=0.229\fwidth,
height=\fheight,
at={(0\fwidth,0\fheight)},
scale only axis,
point meta min=-0.3,
point meta max=1,
axis on top,
xmin=0.4,
xmax=4.4,
xtick={0, 1, 2, 3, 4, 5},
xlabel={width (\SI{}{m})},
ymin=0.4,
ymax=5.4,
ytick={0, 1, 2, 3, 4, 5},
ylabel={depth (\SI{}{m})},
axis background/.style={fill=white},
major tick length = .5em, ylabel style={at={(axis description cs:0.026,0.5)}, xshift=.0em, anchor=north}, yticklabel style={rotate=90}, xlabel style={yshift=.016\fheight, xshift=0}, title style={at={(axis description cs:0.1,0.9)}, yshift=-1.1em, anchor=south, font=\normalfont}
]
\addplot [forget plot] graphics [xmin=0.383333333333333,xmax=4.41666666666667,ymin=0.383333333333333,ymax=5.41666666666667] {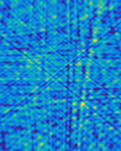};
\addplot [color=mycolor1,only marks,mark size=3.0pt,mark=triangle,mark options={solid,rotate=180,draw=black},forget plot]
  table[row sep=crcr]{%
3.13457459214321	1.9645535740877\\
}; \label{mapExamples_nf_0} 
\addplot [color=mycolor2,only marks,mark size=3.0pt,mark=triangle,mark options={solid,draw=black},forget plot]
  table[row sep=crcr]{%
3.13333333333333	1.96666666666667\\
}; \label{mapExamples_nf_1} 
\addplot [color=mycolor3,only marks,mark size=2.0pt,mark=*,mark options={solid,fill=white,draw=black},forget plot]
  table[row sep=crcr]{%
0.4	0.4\\
0.4	5.4\\
4.4	5.4\\
4.4	0.4\\
}; \label{mapExamples_nf_2} 
\node[fill=white, right, align=left, text=black]
at (axis cs:1,4.5) {\scriptsize conv. FD-SRP};
\end{axis}

\begin{axis}[%
width=0.229\fwidth,
height=\fheight,
at={(0.245\fwidth,0\fheight)},
scale only axis,
point meta min=-0.3,
point meta max=1,
axis on top,
xmin=0.4,
xmax=4.4,
xtick={0, 1, 2, 3, 4, 5},
xlabel={width (\SI{}{m})},
ymin=0.4,
ymax=5.4,
ytick={\empty},
axis background/.style={fill=white},
major tick length = .5em, ylabel style={at={(axis description cs:0.026,0.5)}, xshift=.0em, anchor=north}, yticklabel style={rotate=90}, xlabel style={yshift=.016\fheight, xshift=0}, title style={at={(axis description cs:0.1,0.9)}, yshift=-1.1em, anchor=south, font=\normalfont}
]
\addplot [forget plot] graphics [xmin=0.383333333333333,xmax=4.41666666666667,ymin=0.383333333333333,ymax=5.41666666666667] {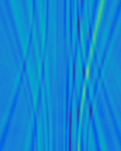};
\addplot [color=mycolor1,solid,mark size=3.0pt,mark=triangle,mark options={solid,rotate=180,draw=black},forget plot]
  table[row sep=crcr]{%
3.13457459214321	1.9645535740877\\
}; \label{mapExamples_nf_3} 
\addplot [color=mycolor2,solid,mark size=3.0pt,mark=triangle,mark options={solid,draw=black},forget plot]
  table[row sep=crcr]{%
3.3	3\\
}; \label{mapExamples_nf_4} 
\addplot [color=mycolor3,solid,mark size=2.0pt,mark=*,mark options={solid,fill=white,draw=black},forget plot]
  table[row sep=crcr]{%
0.4	0.4\\
0.4	5.4\\
4.4	5.4\\
4.4	0.4\\
}; \label{mapExamples_nf_5} 
\node[fill=white, right, align=left, text=black]
at (axis cs:1,4.5) {\scriptsize  LR-SRP};
\end{axis}

\begin{axis}[%
width=0.229\fwidth,
height=\fheight,
at={(0.49\fwidth,0\fheight)},
scale only axis,
point meta min=-0.3,
point meta max=1,
axis on top,
xmin=0.4,
xmax=4.4,
xtick={0, 1, 2, 3, 4, 5},
xlabel={width (\SI{}{m})},
ymin=0.4,
ymax=5.4,
ytick={\empty},
axis background/.style={fill=white},
major tick length = .5em, ylabel style={at={(axis description cs:0.026,0.5)}, xshift=.0em, anchor=north}, yticklabel style={rotate=90}, xlabel style={yshift=.016\fheight, xshift=0}, title style={at={(axis description cs:0.1,0.9)}, yshift=-1.1em, anchor=south, font=\normalfont}
]
\addplot [forget plot] graphics [xmin=0.383333333333333,xmax=4.41666666666667,ymin=0.383333333333333,ymax=5.41666666666667] {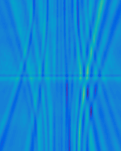};
\addplot [color=mycolor1,solid,mark size=3.0pt,mark=triangle,mark options={solid,rotate=180,draw=black},forget plot]
  table[row sep=crcr]{%
3.13457459214321	1.9645535740877\\
}; \label{mapExamples_nf_6} 
\addplot [color=mycolor2,solid,mark size=3.0pt,mark=triangle,mark options={solid,draw=black},forget plot]
  table[row sep=crcr]{%
3.3	3\\
}; \label{mapExamples_nf_7} 
\addplot [color=mycolor3,solid,mark size=2.0pt,mark=*,mark options={solid,fill=white,draw=black},forget plot]
  table[row sep=crcr]{%
0.4	0.4\\
0.4	5.4\\
4.4	5.4\\
4.4	0.4\\
}; \label{mapExamples_nf_8} 
\node[fill=white, right, align=left, text=black]
at (axis cs:1,4.5) {\scriptsize SLRI-SRP (proposed)};
\end{axis}

\begin{axis}[%
width=0.229\fwidth,
height=\fheight,
at={(0.736\fwidth,0\fheight)},
scale only axis,
point meta min=-0.3,
point meta max=1,
axis on top,
xmin=0.4,
xmax=4.4,
xtick={0, 1, 2, 3, 4, 5},
xlabel={width (\SI{}{m})},
ymin=0.4,
ymax=5.4,
ytick={\empty},
axis background/.style={fill=white},
major tick length = .5em, ylabel style={at={(axis description cs:0.026,0.5)}, xshift=.0em, anchor=north}, yticklabel style={rotate=90}, xlabel style={yshift=.016\fheight, xshift=0}, title style={at={(axis description cs:0.1,0.9)}, yshift=-1.1em, anchor=south, font=\normalfont}
]
\addplot [forget plot] graphics [xmin=0.383333333333333,xmax=4.41666666666667,ymin=0.383333333333333,ymax=5.41666666666667] {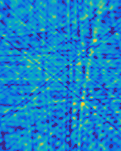};
\addplot [color=mycolor1,solid,mark size=3.0pt,mark=triangle,mark options={solid,rotate=180,draw=black},forget plot]
  table[row sep=crcr]{%
3.13457459214321	1.9645535740877\\
}; \label{mapExamples_nf_9} 
\addplot [color=mycolor2,solid,mark size=3.0pt,mark=triangle,mark options={solid,draw=black},forget plot]
  table[row sep=crcr]{%
3.13333333333333	1.96666666666667\\
}; \label{mapExamples_nf_10} 
\addplot [color=mycolor3,solid,mark size=2.0pt,mark=*,mark options={solid,fill=white,draw=black},forget plot]
  table[row sep=crcr]{%
0.4	0.4\\
0.4	5.4\\
4.4	5.4\\
4.4	0.4\\
}; \label{mapExamples_nf_11} 
\node[fill=white, right, align=left, text=black]
at (axis cs:1,4.5) {\scriptsize SSPI-SRP (proposed)};
\end{axis}

\begin{axis}[%
width=0.019\fwidth,
height=\fheight,
at={(0.981\fwidth,0\fheight)},
scale only axis,
point meta min=-0.3,
point meta max=1,
axis on top,
xmin=0,
xmax=1,
xtick={\empty},
ymin=-0.3,
ymax=1,
ytick={  0, 0.5,   1},
axis background/.style={fill=white},
axis x line*=bottom,
axis y line*=right,
major tick length = .5em, ylabel style={at={(axis description cs:0.026,0.5)}, xshift=.0em, anchor=north}, yticklabel style={rotate=90}, xlabel style={yshift=.016\fheight, xshift=0}, title style={at={(axis description cs:0.1,0.9)}, yshift=-1.1em, anchor=south, font=\normalfont}
]
\addplot [forget plot] graphics [xmin=-0.5,xmax=1.5,ymin=-0.306565656565657,ymax=1.00656565656566] {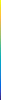};
\end{axis}
\end{tikzpicture}%
\caption{Example map of conv. {FD-}SRP and corresponding maps of LR-SRP, SLRI-SRP (proposed), and SSPI-SRP (proposed) with relative complexity $C_\mathrm{rel} \approx 2\cdot 10^{-2}$  (i.e. nearly two orders of magnitude  less  than conv. {FD-}SRP) in the NF scenario at $\mathit{SNR} = \SI{0}{dB}$ and reverberation time $T_{60} = \SI{0.2}{s}$. The maps represent the horizontal slice of the (three-dimensional) search grid closest to the true source location. The (projected) microphone locations, the true source location, and the estimated source location are marked by [\ref{mapExamples_nf_2}], [\ref{mapExamples_nf_0}], and [\ref{mapExamples_nf_1}], respectively. {The corresponding SI-SRP map with $C_\mathrm{rel} \approx 0.25$ is visually indistinguishable from the conv. SPR map (cf. also the map error in Fig. \ref{fig:nf_matMapErr}) and is left out for brevity.}}. 
\label{fig:nf_map}
\end{figure*}

\setlength\fwidth{7.85cm}
\setlength\fheight{8.5cm} 

 \begin{figure}[t]
\centering
\hspace*{-0.18cm} 
%
%
\definecolor{mycolor1}{rgb}{0.00000,0.44700,0.74100}%
\definecolor{mycolor2}{rgb}{0.85000,0.32500,0.09800}%
\definecolor{mycolor3}{rgb}{0.92900,0.69400,0.12500}%
\begin{tikzpicture}

\begin{axis}[%
width=\fwidth,
height=0.215\fheight,
at={(0\fwidth,0.725\fheight)},
scale only axis,
point meta min=-0.3,
point meta max=1,
axis on top,
xmin=0,
xmax=358,
xtick={0,60,120,180,240,300,360},
xticklabels={\empty},
ymin=90,
ymax=180,
ytick={ 90, 150},
ylabel={pol. angle ($^\circ$)},
axis background/.style={fill=white},
major tick length = .5em, ylabel style={at={(axis description cs:0.026,0.5)}, xshift=.0em, anchor=north}, yticklabel style={rotate=90}, xlabel style={at={(ticklabel cs: -0.01,-8)}, anchor=west}, title style={at={(axis description cs:0.1,0.9)}, yshift=-1.1em, anchor=south, font=\normalfont}
]
\addplot [forget plot] graphics [xmin=-1,xmax=359,ymin=89,ymax=181] {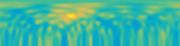};
\addplot [color=mycolor1,solid,mark size=3.0pt,mark=triangle,mark options={solid,rotate=180,draw=black},forget plot]
  table[row sep=crcr]{%
140.343121671246	148.535648630837\\
}; \label{mapExamples_ff_0} 
\addplot [color=mycolor2,solid,mark size=3.0pt,mark=triangle,mark options={solid,draw=black},forget plot]
  table[row sep=crcr]{%
142	150\\
}; \label{mapExamples_ff_1} 
\addplot [color=mycolor3,solid,mark size=2.0pt,mark=*,mark options={solid,fill=white,draw=black},forget plot]
  table[row sep=crcr]{%
0	90\\
60	90\\
120	90\\
180	90\\
240	90\\
300	90\\
}; \label{mapExamples_ff_2} 
\node[fill=white, right, align=left, text=black]
at (axis cs:225,157.5) {\scriptsize conv. FD-SRP};
\end{axis}

\begin{axis}[%
width=\fwidth,
height=0.215\fheight,
at={(0\fwidth,0.483\fheight)},
scale only axis,
point meta min=-0.3,
point meta max=1,
axis on top,
xmin=0,
xmax=358,
xtick={0,60,120,180,240,300,360},
xticklabels={\empty},
ymin=90,
ymax=180,
ytick={ 90, 150},
ylabel={pol. angle ($^\circ$)},
axis background/.style={fill=white},
major tick length = .5em, ylabel style={at={(axis description cs:0.026,0.5)}, xshift=.0em, anchor=north}, yticklabel style={rotate=90}, xlabel style={at={(ticklabel cs: -0.01,-8)}, anchor=west}, title style={at={(axis description cs:0.1,0.9)}, yshift=-1.1em, anchor=south, font=\normalfont}
]
\addplot [forget plot] graphics [xmin=-1,xmax=359,ymin=89,ymax=181] {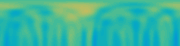};
\addplot [color=mycolor1,solid,mark size=3.0pt,mark=triangle,mark options={solid,rotate=180,draw=black},forget plot]
  table[row sep=crcr]{%
140.343121671246	148.535648630837\\
}; \label{mapExamples_ff_3} 
\addplot [color=mycolor2,solid,mark size=3.0pt,mark=triangle,mark options={solid,draw=black},forget plot]
  table[row sep=crcr]{%
170	160\\
}; \label{mapExamples_ff_4} 
\addplot [color=mycolor3,solid,mark size=2.0pt,mark=*,mark options={solid,fill=white,draw=black},forget plot]
  table[row sep=crcr]{%
0	90\\
60	90\\
120	90\\
180	90\\
240	90\\
300	90\\
}; \label{mapExamples_ff_5} 
\node[fill=white, right, align=left, text=black]
at (axis cs:225,157.5) {\scriptsize LR-SRP};
\end{axis}

\begin{axis}[%
width=\fwidth,
height=0.215\fheight,
at={(0\fwidth,0.242\fheight)},
scale only axis,
point meta min=-0.3,
point meta max=1,
axis on top,
xmin=0,
xmax=358,
xtick={0,60,120,180,240,300,360},
xticklabels={\empty},
ymin=90,
ymax=180,
ytick={ 90, 150},
ylabel={pol. angle ($^\circ$)},
axis background/.style={fill=white},
major tick length = .5em, ylabel style={at={(axis description cs:0.026,0.5)}, xshift=.0em, anchor=north}, yticklabel style={rotate=90}, xlabel style={at={(ticklabel cs: -0.01,-8)}, anchor=west}, title style={at={(axis description cs:0.1,0.9)}, yshift=-1.1em, anchor=south, font=\normalfont}
]
\addplot [forget plot] graphics [xmin=-1,xmax=359,ymin=89,ymax=181] {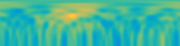};
\addplot [color=mycolor1,solid,mark size=3.0pt,mark=triangle,mark options={solid,rotate=180,draw=black},forget plot]
  table[row sep=crcr]{%
140.343121671246	148.535648630837\\
}; \label{mapExamples_ff_6} 
\addplot [color=mycolor2,solid,mark size=3.0pt,mark=triangle,mark options={solid,draw=black},forget plot]
  table[row sep=crcr]{%
144	150\\
}; \label{mapExamples_ff_7} 
\addplot [color=mycolor3,solid,mark size=2.0pt,mark=*,mark options={solid,fill=white,draw=black},forget plot]
  table[row sep=crcr]{%
0	90\\
60	90\\
120	90\\
180	90\\
240	90\\
300	90\\
}; \label{mapExamples_ff_8} 
\node[fill=white, right, align=left, text=black]
at (axis cs:225,157.5) {\scriptsize SLRI-SRP (proposed)};
\end{axis}

\begin{axis}[%
width=\fwidth,
height=0.215\fheight,
at={(0\fwidth,0\fheight)},
scale only axis,
point meta min=-0.3,
point meta max=1,
axis on top,
xmin=0,
xmax=358,
xtick={  0,  60, 120, 180, 240, 300, 360},
xlabel={azimuth angle ($^\circ$)},
ymin=90,
ymax=180,
ytick={ 90, 150},
ylabel={pol. angle ($^\circ$)},
axis background/.style={fill=white},
major tick length = .5em, ylabel style={at={(axis description cs:0.026,0.5)}, xshift=.0em, anchor=north}, yticklabel style={rotate=90}, xlabel style={at={(ticklabel cs: -0.01,-8)}, anchor=west}, title style={at={(axis description cs:0.1,0.9)}, yshift=-1.1em, anchor=south, font=\normalfont}
]
\addplot [forget plot] graphics [xmin=-1,xmax=359,ymin=89,ymax=181] {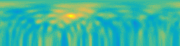};
\addplot [color=mycolor1,solid,mark size=3.0pt,mark=triangle,mark options={solid,rotate=180,draw=black},forget plot]
  table[row sep=crcr]{%
140.343121671246	148.535648630837\\
}; \label{mapExamples_ff_9} 
\addplot [color=mycolor2,solid,mark size=3.0pt,mark=triangle,mark options={solid,draw=black},forget plot]
  table[row sep=crcr]{%
140	150\\
}; \label{mapExamples_ff_10} 
\addplot [color=mycolor3,solid,mark size=2.0pt,mark=*,mark options={solid,fill=white,draw=black},forget plot]
  table[row sep=crcr]{%
0	90\\
60	90\\
120	90\\
180	90\\
240	90\\
300	90\\
}; \label{mapExamples_ff_11} 
\node[fill=white, right, align=left, text=black]
at (axis cs:225,157.5) {\scriptsize SSPI-SRP (proposed)};
\end{axis}

\begin{axis}[%
width=\fwidth,
height=0.034\fheight,
at={(0\fwidth,0.966\fheight)},
scale only axis,
point meta min=-0.3,
point meta max=1,
axis on top,
xmin=-0.3,
xmax=1,
xtick={  0, 0.5,   1},
ymin=0,
ymax=0.0630918228683592,
ytick={\empty},
axis background/.style={fill=white},
axis x line*=top,
axis y line*=left,
major tick length = .5em, ylabel style={at={(axis description cs:0.026,0.5)}, xshift=.0em, anchor=north}, yticklabel style={rotate=90}, xlabel style={at={(ticklabel cs: -0.01,-8)}, anchor=west}, title style={at={(axis description cs:0.1,0.9)}, yshift=-1.1em, anchor=south, font=\normalfont}
]
\addplot [forget plot] graphics [xmin=-0.306565656565657,xmax=1.00656565656566,ymin=-0.0315459114341796,ymax=1.03154591143418] {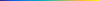};
\end{axis}
\end{tikzpicture}%
\caption{Example map of conv. {FD-}SRP and corresponding maps of LR-SRP, SLRI-SRP (proposed), and SSPI-SRP (proposed) with relative complexity $C_\mathrm{rel} \approx 1 \cdot 10^{-2}$ (i.e. two orders of magnitude less than conv. {FD-}SRP) in the FF scenario with array diameter $\diameter=\SI{30}{cm}$ at $\mathit{SNR} = \SI{0}{dB}$ and reverberation time $T_{60} = \SI{0.6}{s}$. The microphone locations, the true source location, and the estimated source location are marked by [\ref{mapExamples_nf_2}], [\ref{mapExamples_nf_0}], and [\ref{mapExamples_nf_1}], respectively. {The corresponding SI-SRP map with $C_\mathrm{rel} \approx 3 \cdot 10^{-2}$ is visually indistinguishable from the conv. SPR map (cf. also the map error in Fig. \ref{fig:ff_matMapErr}) and is left out for brevity.}}. 
\label{fig:ff_map}
\end{figure}

\begin{table}[t]
	\caption{Parameters relevant to computational complexity in the NF and FF scenario. In LR-SRP, SLRI-SRP, and SSPI-SRP,  $C_{\mathrm{rel}}$ is scalable and varied by means of the low-rank or sparsity constraint.}
	\begin{center}
		\begin{tabular}{@{} l l l l} 
			\toprule
			parameter\textbackslash scenario & & NF & FF \\
			\midrule
			grid resolution &  & \SI{3.33}{cm} &  2$^\circ$ \\
		    \# candidate locations & $J$ & $73084$ & $8101$ \\
			\# microphones & $M$ & $4$ & $6$ \\
			\# microphone pairs & $P$ & $6$ & $15$ \\
			largest microphone distance & $d_p$ &  \SI{6.4}{m} & \SI{10}{cm}--\SI{40}{cm}\\
			twice the largest TDOA limit & $2\Delta t_{p|0}$ & \SI{37.7}{ms} & \SI{3.5}{ms}\\
		    frame duration & $2KT$ & \SI{128}{ms} & \SI{64}{ms}\\
			sampling rate & $1/T$ & \SI{4}{kHz} & \SI{16}{kHz}\\
			number of frequency bins & $K-1$ & $255$ & $511$ \\
			mean \# required samples & $N$ & $125$ & $11.4$--$32.2$ \\
		    complexity of conv. {FD-}SRP & $C$ & $111.8\cdot 10^6$ & $62.1\cdot 10^6$ \\
			rel. complexity of SI-SRP & $C_{\mathrm{rel}}$ & $0.25$ & $0.016$--$0.037$ \\
			\bottomrule
		\end{tabular}
	\end{center}
   \label{tab:simparameters}
\end{table} 

\begin{table}[t]
	\caption{DRR and critical distance for the simulated NF and FF scenario at given reverberation times.}
	\begin{center}
		\begin{tabular}{@{} c | c c | c c} 
			\toprule
			 reverb. time $T_{60}$ (\SI{}{s}) & \multicolumn{2}{c |}{mean $\mathit{DRR}$ (\SI{}{dB})}  & \multicolumn{2}{c}{critical dist. (\SI{}{m})} \\
			\midrule
			 NF and FF & NF & FF & NF & FF \\
			 \midrule
			 $0$ & $\infty$ & $\infty$ & $\infty$& $\infty$\\
			 $0.2$ & $-6.3$ & $-2.6$ & $1.28$&$2.07$\\
			 $0.6$ & $-12.2$& $-8.8$ & $0.74$&$1.19$\\
			\bottomrule
		\end{tabular}
	\end{center}
   \label{tab:reverbparameters}
\end{table} 

\subsection{Algorithms and Performance Measures}
\label{sec:perm}

Next to the various SRP approaches discussed in Sec. \ref{sec:SRP} to Sec. \ref{sec:lr_sp_int} yielding a fine SRP map, we also evaluate the localization accuracy of the non-exhaustive spatial search approach in \cite{marti13}, which  iteratively refines the search space based on a accumulating TD GCC samples over volumes of decreasing dimensions. In the remainder, we refer to this volumetric approach as iterative refinement-based SRP (IR-SRP) and denote it by the subscript $_{\mathrm{ir}}$. 
The complexity of IR-SRP is defined as the sum of $C_{\mathrm{samp}}$ and the number of {MAC operations} required for accumulating the TD GCC samples over all refinement steps \cite{marti13}.

We define the following performance measures.
For $C_{\mathrm{type}} \in \{C_{\mathrm{td}},\, C_{\mathrm{lr}},\,C_{\mathrm{si}},\,C_{\mathrm{slri}},\,C_{\mathrm{sspi}}, \,C_{\mathrm{ir}}\}$, we define the relative complexity with respect to $C$ as
\begin{align}
C_{\mathrm{rel}} &= C_{\mathrm{type}}/C.
\end{align}
The SRP matrix error $\varepsilon_H(\mathbf{H}_{{\operatorname{type}}})$ with respect to $\mathbf{H}$ for  $ \mathbf{H}_{{\operatorname{type}}} \in \{\mathbf{H}_{\mathrm{lr}},\,\mathbf{H}_{\mathrm{si}},\,\mathbf{H}_{\mathrm{slri}},\,\mathbf{H}_{\mathrm{sspi}}\}$ and the SRP map error $\varepsilon_z(\mathbf{z}_{\mathrm{type}})$ with respect to $\mathbf{z}$ for $\mathbf{z}_{\mathrm{type}}  \in \{\mathbf{z}_{\mathrm{lr}},\,\mathbf{z}_{\mathrm{si}},\,\mathbf{z}_{\mathrm{slri}},\,\mathbf{z}_{\mathrm{sspi}}\}$ are defined as
\begin{align}
\varepsilon_H &= 10\log_{10}
\frac{\displaystyle \Vert\mathbf{H}_{{\operatorname{type}}} - \mathbf{H}\Vert_{\mathrm{F}}^2}{\displaystyle \Vert \mathbf{H}\Vert_{\mathrm{F}}^2}\,\si{dB},\\
\varepsilon_z &= 10\log_{10} \frac{ \Vert \mathbf{z}_{\mathrm{type}} - \mathbf{z}\Vert^2}{\Vert \mathbf{z} \Vert^2}\,\si{dB}.
\end{align}
{The performance in terms of $\varepsilon_H(\mathbf{H}_{{\operatorname{type}}})$ and $\varepsilon_z(\mathbf{z}_{\mathrm{type}})$ versus $C_{\mathrm{rel}}$ is evaluated in Sec. \ref{sec:approxErr}.} 

Further, for the NF or FF source-location estimates $\hat{\mathbf{q}}_{\mathrm{s}|\operatorname{type}} \in \{\hat{\mathbf{q}}_{\mathrm{s}},\,\hat{\mathbf{q}}_{\mathrm{s}|\mathrm{td}},\,\hat{\mathbf{q}}_{\mathrm{s}|\mathrm{lr}},\,\hat{\mathbf{q}}_{\mathrm{s}|\mathrm{si}},\,\hat{\mathbf{q}}_{\mathrm{s}|\mathrm{slri}},\,\hat{\mathbf{q}}_{\mathrm{s}|\mathrm{sspi}}\,,\hat{\mathbf{q}}_{\mathrm{s}|\mathrm{ir}}\}$, we define the localization error $\varepsilon_{\mathrm{s}}$ as the distance between the estimated and the true source location and the angle between the estimated and the true incident direction, respectively, i.e.
\begin{align}
\varepsilon_{\mathrm{s}} &= 
\begin{cases}
\Vert\hat{\mathbf{q}}_{\mathrm{s}|\operatorname{type}}-\mathbf{q}_{s}\Vert &\!\text{for NF localization},\\
\operatorname{acos}[\hat{\mathbf{q}}_{\mathrm{s}|\operatorname{type}}^\transp\mathbf{q}_{\mathrm{s}}] &\!\text{for FF localization}.
\label{eq:err_s}
\end{cases}
\end{align}
As the distributions of $\varepsilon_{\mathrm{s}}$ were observed to have a nearly uniform tail, suggesting that erroneous location estimates are uniformly distributed over the search grid, cf. also Sec. \ref{sec:appearanceSRP}, we further define the localization accuracy $\rho_{\mathrm{s}}$ as a function of $\varepsilon_{\mathrm{s}}$ by
{
\begin{align}
\rho_{\mathrm{s}} = 
\begin{cases}
1  & \text{for } \varepsilon_{\mathrm{s}} \leq \varepsilon_{\mathrm{th}}\\
0 & \text{for } \varepsilon_{\mathrm{s}} > \varepsilon_{\mathrm{th}}
\end{cases},
\end{align}
where we use the NF and FF error tolerances {$\varepsilon_{\mathrm{th}} = \{4,\,8,\,16\}\SI{}{cm}$} and $\varepsilon_{\mathrm{th}} = 3^\circ$, respectively.}
{The performance in terms of $\rho_{\mathrm{s}}$ versus $C_{\mathrm{rel}}$ is evaluated for varying values of $\varepsilon_{\mathrm{th}}$ in the NF case and different array apertures in the FF case in Sec. \ref{sec:locAccThreshAperture}, for for varying noise and reverberation values in \ref{sec:locAcc}, and for varying number of microphones in \ref{sec:resultsLocVarMic}.} 

{Finally, we define $D$ as the computation time of conv. FD SRP, and $D_{\mathrm{type}} \in \{D_{\mathrm{td}},\,D_{\mathrm{lr}},\,D_{\mathrm{si}},\,D_{\mathrm{slri}},\,D_{\mathrm{sspi}},\,D_{\mathrm{ir}}\}$ as the computation time of the other considered SRP approaches.
Based on this, we define the relative computation times with respect to $D$ as
\begin{align}
D_{\mathrm{rel}} &= D_{\mathrm{type}}/D.
\end{align}
The computation time of the iFFTs will be denoted as $D_\mathrm{samp}$.
All computation times are obtained for the implementation in MATLAB \cite{taslp23Code} on a MacBook Pro model with an Intel Core i7 processor.
The performance in terms of $D_{\mathrm{rel}}$ versus $C_{\mathrm{rel}}$ is evaluated in \ref{sec:resultsComputeTime}.
}

\setlength\fwidth{7.9cm}
\setlength\fheight{6cm} 
 \begin{figure}
\centering
\hspace*{-0.285cm} 
    \input{./fig/matrMapErr_nf.tex} 
\caption{Matrix error $\varepsilon_{H}$ (top) and median map error $\varepsilon_{z}$ (bottom) vs. relative complexity $C_{\mathrm{rel}}$ for LR-SRP [\ref{locErr_nf_1}], SI-SRP (single point) [\ref{locErr_nf_8}], SLRI-SRP (proposed) [\ref{locErr_nf_3}], and SSPI-SRP (proposed) [\ref{locErr_nf_6}] in the NF scenario. The shaded areas in the map error indicate the range from the 10$^{\text{th}}$ to the 90$^{\text{th}}$ percentile. The gray area indicates the range where $C_{\mathrm{rel}} < C_\mathrm{samp}$, which cannot be reached by SLRI-SRP and SSPI-SRP.}.
\label{fig:nf_matMapErr}
\end{figure}

\subsection{{Simulations based on Simulated RIRs}}
\label{sec:simRIR}
{To be able to randomize the source location and study the effect of different reverberation times {and array apertures}, we first perform simulations based on simulated RIRs.}

\subsubsection{Simulation Setup}
\label{sec:simsetup}

{To demonstrate the performance of the proposed approaches for both small and large array sizes, we simulate an NF and an FF scenario within shoe-box shaped rooms, respectively.}
The geometries of the simulated scenarios are illustrated in Fig. \ref{fig:nf_geom} and Fig. \ref{fig:ff_geom}, and the parameters relevant to computational complexity are summarized in Table \ref{tab:simparameters}.

In the NF scenario, $M = 4$ microphones are placed near the corners of a room with dimensions $(4.9 \times 5.9 \times 3.5) \SI{}{m}$
at a height of \SI{3}{m} and in a distance of \SI{0.4}{m} to \SI{0.5}{m} to the walls.
The three-dimensional search grid samples a volume of $(4 \times 5 \times 0.1) \SI{}{m}$ and lies below the microphones around a height of \SI{1.5}{m}.
Its spatial resolution is \SI{3.33}{cm}, yielding $J = 73084$ candidate locations.
{Note that while a wider height range will commonly be required in practical applications, we here limit the range to keep simulations more tractable.}
In the FF scenario, a circular microphone array with diameter {$\diameter = \{10,\, 20,\, 30,\, 40\}\SI{}{cm}$} and $M = 6$ microphones is placed in the horizontal plane $\SI{10}{cm}$ below the ceiling and near the center of a room with dimensions $(6.5 \times 9 \times 4.5) \SI{}{m}$.
The two-dimensional search grid samples the polar and azimuth angle of the candidate locations and covers the lower half-sphere with zenith and azimuth direction along the third (i.e., the vertical) and the second room dimension. 
The angular resolution is 2$^\circ$, yielding $J = 8101$ candidate locations.

In both scenarios, a single source is placed in $512$ randomly selected locations, which are not restricted to be on the search grid.
In the NF scenario, the random locations are restricted to be inside the volume spanned by the search grid.
In the FF scenario, the distance of the source to the center of the microphone array is restricted to be at least $\SI{2}{m}$.

{In each source location, speech of 3 female and 3 male speakers \cite{speechVCTK}  of in total 241$\SI{}{s}$ is used as a source signal.} 
Diffuse babble noise \cite{habets2008generating, Dietzen2023} is added at varying SNRs of $\mathit{SNR} = \{12,\,6,\,0,\,-6, \,-12\}\SI{}{dB}$. 
The shoe-box rooms are simulated using the randomized image method \cite{desena15} at varying reverberation times $T_{60} = \{0,\,0.2,\,0.6\}\SI{}{s}$.
Note that due to different geometries, the same reverberation time results in different direct-to-reverberant ratios (DRRs) and critical distances \cite{kuttruff14} across the two scenarios, with the NF scenario being more adversely affected as shown in Table \ref{tab:reverbparameters}.

The speed of sound is set to $c = \SI{340}{m/s}$, and the sampling rates of the NF and the FF scenario are respectively set to $\SI{4}{kHz}$  and $\SI{16}{kHz}$, implying the bandlimits $\nicefrac{\displaystyle \omega_{\mathrm{b}}}{\displaystyle 2\pi} = \SI{2}{kHz}$\footnote{{Note that the required spatial resolution of the search grid depends on the bandlimit of the signal \cite{salvati2017exploiting, GarciaBarrios2021}. In our simulations, we have chosen the spatial resolution and the bandlimit such that the main lobes of the resulting TD GCCs $\xi_{p}(\tau)$ are sampled by at least one candidate location TDOA $\Delta t_{p}(i)$, cf. (\ref{SRP_T_IFT}).} While the bandlimit of speech signals is larger than \SI{2}{kHz}, we low-pass filter the signals accordingly in the NF scenario.
} and $\nicefrac{\displaystyle \omega_{\mathrm{b}}}{\displaystyle 2\pi} = \SI{8}{kHz}$.

\begin{figure}
\centering
\hspace*{-0.285cm} 
    \input{./fig/matrMapErr_ff_r15.tex} 
\caption{Matrix error $\varepsilon_{H}$ (top) and median map error $\varepsilon_{z}$ (bottom) vs. relative complexity $C_{\mathrm{rel}}$ for LR-SRP [\ref{locErr_nf_1}], SI-SRP (single point) [\ref{locErr_nf_8}], SLRI-SRP (proposed) [\ref{locErr_nf_3}], and SSPI-SRP (proposed) [\ref{locErr_nf_6}] in the FF scenario in the FF scenario with array diameter $\diameter=\SI{30}{cm}$. The shaded areas in the map error indicate the range from the 10$^{\text{th}}$ to the 90$^{\text{th}}$ percentile. The gray area indicates the range where $C_{\mathrm{rel}} < C_\mathrm{samp}$, which cannot be reached by SLRI-SRP and SSPI-SRP.}.
\label{fig:ff_matMapErr}
\end{figure}

\setlength\fwidth{13cm}
\setlength\fheight{7cm} 
 \begin{figure*}[t]
\centering
\hspace*{-0.35cm} 
    \input{./fig/locErr_varTheshRadius.tex} 
\caption{
{Mean localization accuracy $\rho_{\mathrm{s}}$ vs. relative complexity $C_{\mathrm{rel}}$ of conv. TD-SRP (single point) [\ref{locErr_nf_10}], LR-SRP [\ref{locErr_nf_1}], SI-SRP (single point) [\ref{locErr_nf_8}], SLRI-SRP (proposed) [\ref{locErr_nf_3}], SSPI-SRP (proposed) [\ref{locErr_nf_6}], {and IR-SRP (single point) [}\ref{locErr_nf_9}{]}  as compared to $\rho_{\mathrm{s}}$ of conv. FD-SRP [\ref{locErr_nf_0}] for different error tolerances $\varepsilon_{\operatorname{th}}$ and diameters $\diameter$ in the NF and FF scenario, respectively,  at $\mathit{SNR} = \SI{12}{dB}$ and $T_{60} = \SI{0}{s}$.
From left to right, the top and the bottom row respectively illustrate the effect of increasing  $\varepsilon_{\operatorname{th}}$  and $\diameter$ in the NF and the FF scenario, respectively. 
The gray area indicates the range where $C_{\mathrm{rel}} < C_\mathrm{samp}$, which cannot be reached by conv. TD-SRP, SLRI-SRP, SSPI-SRP,  and IR-SRP. In the top and the bottom row, the vertical dotted grid line indicates $C_{\mathrm{rel}}$ of SSPI-SRP for $Q_{\Lambda} = JP$ and $Q_{\Lambda} = 5JP$, respectively, i.e. using on average one and five sample(s) per candidate location and microphone pair for interpolation.
}}. 
\label{fig:nf_ff_threshRadius}
\end{figure*}

\setlength\fwidth{13cm}
\setlength\fheight{7cm} 
 \begin{figure*}[t]
\centering
\hspace*{-0.35cm} 
    \input{./fig/locErr_nf.tex} 
\caption{Mean localization accuracy $\rho_{\mathrm{s}}$ vs. relative complexity $C_{\mathrm{rel}}$ of {conv. TD-SRP (single point) [\ref{locErr_nf_10}],} LR-SRP [\ref{locErr_nf_1}], SI-SRP (single point) [\ref{locErr_nf_8}], SLRI-SRP (proposed) [\ref{locErr_nf_3}], SSPI-SRP (proposed) [\ref{locErr_nf_6}], {and IR-SRP (single point) [}\ref{locErr_nf_9}{]}  as compared to $\rho_{\mathrm{s}}$ of conv. {FD-}SRP [\ref{locErr_nf_0}] for different $\mathit{SNR}$ values and reverberation times $T_{60}$ in the NF scenario with error tolerance $\varepsilon_{\mathrm{th}} = \SI{16}{cm}$. From left to right, the top and the bottom row respectively illustrate the effect of decreasing  $\mathit{SNR}$ and increasing  $T_{60}$. 
The gray area indicates the range where $C_{\mathrm{rel}} < C_\mathrm{samp}$, which cannot be reached by conv. TD-SRP, SLRI-SRP, SSPI-SRP,  and IR-SRP. The vertical dotted grid line indicates $C_{\mathrm{rel}}$ of SSPI-SRP for $Q_{\Lambda} = JP$, i.e. using on average one sample per candidate location and microphone pair for interpolation {and corresponding to $C_{\mathrm{rel}}$ of  conv. TD-SRP.}}
\label{fig:nf_locErr}
\end{figure*}

\setlength\fwidth{13cm}
\setlength\fheight{7cm} 
 \begin{figure*}[t]
\centering
\hspace*{-0.35cm} 
    \input{./fig/locErr_ff.tex} 
\caption{Mean localization accuracy $\rho_{\mathrm{s}}$ vs. relative complexity $C_{\mathrm{rel}}$ of {conv. TD-SRP (single point) [\ref{locErr_nf_10}],} LR-SRP [\ref{locErr_nf_1}], SI-SRP (single point) [\ref{locErr_nf_8}], SLRI-SRP (proposed) [\ref{locErr_nf_3}], SSPI-SRP (proposed) [\ref{locErr_nf_6}], {and IR-SRP (single point) [}\ref{locErr_nf_9}{]}  as compared to $\rho_{\mathrm{s}}$ of conv. {FD-}SRP [\ref{locErr_nf_0}] for different $\mathit{SNR}$ values and reverberation times $T_{60}$ iin the FF scenario with array diameter $\diameter=\SI{30}{cm}$. From left to right, the top and the bottom row respectively illustrate the effect of decreasing  $\mathit{SNR}$ and increasing  $T_{60}$. 
The gray area indicates the range where $C_{\mathrm{rel}} < C_\mathrm{samp}$, which cannot be reached by conv. TD-SRP, SLRI-SRP, SSPI-SRP,  and IR-SRP. The vertical dotted grid line indicates $C_{\mathrm{rel}}$ of SSPI-SRP for $Q_{\Lambda} = 5JP$, i.e. using on average one sample per candidate location and microphone pair for interpolation.}. 
\label{fig:ff_locErr}
\end{figure*}

\setlength\fwidth{3.8cm}
\setlength\fheight{.87\fwidth} 
 \begin{figure}
\centering
\hspace*{-0.5cm} 
%
%
\definecolor{mycolor1}{rgb}{0.00000,0.44700,0.74100}%
\begin{tikzpicture}

\begin{axis}[%
width=0.951\fwidth,
height=\fheight,
at={(0\fwidth,0\fheight)},
scale only axis,
xmin=-0.65,
xmax=0.75,
xtick={-0.5,    0,  0.5},
xlabel={width ($\SI{}{m}$)},
ymin=-0.65,
ymax=0.65,
ytick={-0.5,    0,  0.5},
ylabel={depth ($\SI{}{m}$)},
axis background/.style={fill=white},
major tick length = .5em, ylabel style={at={(axis description cs:0.026,0.5)}, xshift=.0em, anchor=north}, xlabel style={yshift=.02\fheight, xshift=0}, yticklabel style={rotate=90}, title style={at={(axis description cs:0.1,0.9)}, yshift=-1.1em, anchor=south, font=\normalfont}
]
\addplot [color=mycolor1,mark size=2.0pt,only marks,mark=*,mark options={solid,fill=white,draw=black},forget plot]
  table[row sep=crcr]{%
-0.5	-0.5\\
0.5	-0.5\\
0.5	0.5\\
-0.5	0.5\\
0	0\\
0	-0.5\\
0.5	0\\
0	0.5\\
-0.5	0\\
-0.25	-0.25\\
0.25	-0.25\\
0.25	0.25\\
-0.25	0.25\\
}; \label{micsRealRIR_0} 
\node[right, align=left, text=black]
at (axis cs:-0.47,-0.48) {\scriptsize $1$};
\node[right, align=left, text=black]
at (axis cs:0.53,-0.48) {\scriptsize $2$};
\node[right, align=left, text=black]
at (axis cs:0.53,0.52) {\scriptsize $3$};
\node[right, align=left, text=black]
at (axis cs:-0.47,0.52) {\scriptsize $4$};
\node[right, align=left, text=black]
at (axis cs:0.03,0.02) {\scriptsize $5$};
\node[right, align=left, text=black]
at (axis cs:0.03,-0.48) {\scriptsize $6$};
\node[right, align=left, text=black]
at (axis cs:0.53,0.02) {\scriptsize $7$};
\node[right, align=left, text=black]
at (axis cs:0.03,0.52) {\scriptsize $8$};
\node[right, align=left, text=black]
at (axis cs:-0.47,0.02) {\scriptsize $9$};
\node[right, align=left, text=black]
at (axis cs:-0.22,-0.23) {\scriptsize $10$};
\node[right, align=left, text=black]
at (axis cs:0.28,-0.23) {\scriptsize $11$};
\node[right, align=left, text=black]
at (axis cs:0.28,0.27) {\scriptsize $12$};
\node[right, align=left, text=black]
at (axis cs:-0.22,0.27) {\scriptsize $13$};
\end{axis}
\end{tikzpicture}%
\caption{{Microphone locations [}\ref{micsRealRIR_0}{] and microphone indeces $m$ in the NF scenario with measured RIRs.}}
\label{fig:nfreal_geom}
\end{figure}

\setlength\fwidth{17.1cm}
\setlength\fheight{7cm} 
 \begin{figure*}[t]
\centering
\hspace*{-0.35cm} 
    \input{./fig/locErr_nfreal2.tex} 
\caption{
{Mean localization accuracy $\rho_{\mathrm{s}}$ vs. relative complexity $C_{\mathrm{rel}}$ of {conv. TD-SRP (single point) [\ref{locErr_nf_10}],} LR-SRP [}\ref{locErr_nf_1}{], SI-SRP (single point) [}\ref{locErr_nf_8}{], SLRI-SRP (proposed) [}\ref{locErr_nf_3}{], SSPI-SRP (proposed) [}\ref{locErr_nf_6}{], and IR-SRP (single point) [}\ref{locErr_nf_9}{]  as compared to $\rho_{\mathrm{s}}$ of conv. {FD-}SRP [\ref{locErr_nf_0}] for increasing number of microphones $M$ from left to right at $\mathit{SNR} = 0\SI{}{dB}$  in the NF scenario with measured RIRs and error tolerance $\varepsilon_{\mathrm{th}} = \SI{16}{cm}$. The gray area indicates the range where $C_{\mathrm{rel}} < C_\mathrm{samp}$, which cannot be reached by conv. TD-SRP, SLRI-SRP, SSPI-SRP, and IR-SRP. The vertical dotted grid line indicates $C_{\mathrm{rel}}$ of SSPI-SRP for $Q_{\Lambda} = JP$, i.e. using on average one samples per candidate location and microphone pair for interpolation {and corresponding to $C_{\mathrm{rel}}$ of  conv. TD-SRP.}}
} 
\label{fig:nfreal_locErr}
\end{figure*}

\setlength\fwidth{7.9cm}
\setlength\fheight{4.5cm} 
 \begin{figure}
\centering
\hspace*{-0.285cm} 
%
%
\definecolor{mycolor1}{rgb}{0.00000,0.44700,0.74100}%
\definecolor{mycolor2}{rgb}{0.85000,0.32500,0.09800}%
\definecolor{mycolor3}{rgb}{0.90000,0.25000,0.25000}%
\definecolor{mycolor4}{rgb}{0.92900,0.69400,0.12500}%
\definecolor{mycolor5}{rgb}{0.25000,0.25000,0.90000}%
\definecolor{mycolor6}{rgb}{0.49400,0.18400,0.55600}%
\definecolor{mycolor7}{rgb}{0.46600,0.67400,0.18800}%
\definecolor{mycolor8}{rgb}{0.30100,0.74500,0.93300}%
\begin{tikzpicture}

\begin{axis}[%
width=0.951\fwidth,
height=\fheight,
at={(0\fwidth,0\fheight)},
scale only axis,
xmode=log,
xmin=0.001,
xmax=0.2,
xtick={0.0001,  0.001,   0.01,    0.1},
xminorticks=true,
xlabel={relative complexity $C_{\mathrm{rel}}$},
xmajorgrids,
xminorgrids,
ymode=log,
ymin=0.001,
ymax=0.2,
ytick={0.001,0.01,0.1},
yticklabels={{ },{$10^{-2}$},{$10^{-1}$}},
yminorticks=true,
ylabel={relative comp. time $D_{\mathrm{rel}}$},
ymajorgrids,
yminorgrids,
axis background/.style={fill=white},
axis x line*=bottom,
axis y line*=left,
set layers=Bowpark, major tick length = .5em, ylabel style={at={(axis description cs:0.026,0.5)}, xshift=.0em, anchor=north}, yticklabel style={rotate=90}, xlabel style={yshift=.005\fheight, xshift=0},title style={at={(axis description cs:0.1,0.9)}, yshift=-1.1em, anchor=south, font=\normalfont}
]
\draw[solid, fill=white!85!black, draw=none] (axis cs:1e-05,1e-05) rectangle (axis cs:0.0015130403689218,1.00001);
\draw[solid, fill=white!85!black, draw=none] (axis cs:1e-05,1e-05) rectangle (axis cs:1.00001,0.0014940927361422);
\addplot [color=white!85!black,dotted,line width=2.0pt,forget plot]
  table[row sep=crcr]{%
0.00035	0.00035\\
0.35	0.35\\
}; \label{computation_time_0} 
\addplot [color=mycolor1,mark size=3.0pt,only marks,mark=triangle*,mark options={solid,rotate=180,fill=black,draw=white},forget plot]
  table[row sep=crcr]{%
0.005450048242938	0.0130961780606142\\
}; \label{computation_time_1} 
\addplot [color=mycolor2,mark size=2.2pt,only marks,mark=*,mark options={solid,fill=white,draw=mycolor3},forget plot]
  table[row sep=crcr]{%
0.005313331189203	0.00724585531627897\\
}; \label{computation_time_2} 
\addplot [color=mycolor4,mark size=3.0pt,only marks,mark=triangle*,mark options={solid,rotate=90,fill=mycolor5,draw=white},forget plot]
  table[row sep=crcr]{%
0.092501666790619	0.0471070994976806\\
}; \label{computation_time_3} 
\addplot [color=mycolor6,mark size=1.9pt,only marks,mark=square*,mark options={solid,fill=mycolor5,draw=white},forget plot]
  table[row sep=crcr]{%
0.005490086515504	0.00406862760163872\\
}; \label{computation_time_4} 
\addplot [color=mycolor7,mark size=2.2pt,only marks,mark=*,mark options={solid,fill=mycolor5,draw=white},forget plot]
  table[row sep=crcr]{%
0.005450048242938	0.0138646031773239\\
}; \label{computation_time_5} 
\addplot [color=mycolor8,mark size=3.0pt,only marks,mark=diamond*,mark options={solid,fill=black,draw=white},forget plot]
  table[row sep=crcr]{%
0.001558342514842	0.00269287319012267\\
}; \label{computation_time_6} 
\end{axis}
\end{tikzpicture}%
\caption{{Relative computation time $D_{\mathrm{rel}}$ vs. relative complexity $C_{\mathrm{rel}}$ of conv. TD-SRP [}\ref{computation_time_1}{, symbol partially covered], LR-SRP [}\ref{computation_time_2}{], SI-SRP [}\ref{computation_time_3}{], SLRI-SRP (proposed) [}\ref{computation_time_4}{], SSPI-SRP (proposed) [}\ref{computation_time_5}{], and IR-SRP  [}\ref{computation_time_6}{]  for $M = 9$ microphones in the NF scenario with measured RIRs, where $C_{\mathrm{rel}}$ of the scalable approaches SL-SRP, SLRI-SRP, and SSPI-SRP is set to $C_{\mathrm{rel}}\approx 5.3\cdot10^{-3}$, corresponding to $C_{\mathrm{rel}}$ of  conv. TD-SRP.  The gray area indicates the range where $C_{\mathrm{rel}} < C_\mathrm{samp}$ or $D_{\mathrm{rel}} < D_\mathrm{samp}$. The dotted line indicates $D_{\mathrm{rel}} = C_{\mathrm{rel}}$.} }
\label{fig:compTime}
\end{figure}

The microphone signals are processed by a short-time Fourier transform (STFT) using square-root Hann windows of $2K = 512$ samples in the NF scenario and $2K = 1024$ in the FF scenario, corresponding to \SI{128}{ms} and \SI{64}{ms} and well exceeding the possible TDOA range, cf. Table \ref{tab:simparameters}. 
For each random source location, SRP maps are computed for 32 different frames of microphone data. 

As shown in Table \ref{tab:simparameters}, the computational complexity of conv. {FD-}SRP in the NF and FF scenario is $C = 111.8\cdot 10^6$ and  $C = 62.1\cdot 10^6$, respectively.
In conv. TD-SRP, SI-SRP, SLRI-SRP, SSPI-SRP, and IR-SRP, {$\boldsymbol{\xi}$ is computed using the iFFT.
W}e set $N_{\mathrm{aux}}$ in (\ref{eq:N_p}) to $N_{\mathrm{aux}} = 2$, resulting in $N = 125$ and {$N = \{11.4,\, 18.2,\, 25.8,\, 32.2\}\SI{}{cm}$} samples per microphone pair in the NF and FF scenario{, where in the FF scenario larger values correspond to larger array diameters}.
For SI-SRP, we obtain the relative complexities $C_{\mathrm{rel}} = 0.25$ and {$C_{\mathrm{rel}} = \{1.6,\, 2.3,\, 3,\, 3.7\}\cdot10^{-2}$}. 
In LR-SRP, SLRI-SRP, and SSPI-SRP,  $C_{\mathrm{rel}}$ is scalable and varied by means of the low-rank or sparsity constraint.
In IR-SRP, the spatial resolution is iteratively increased from \SI{40}{cm} to \SI{13.33}{cm} to \SI{3.33}{cm} in the NF scenario, and from 12$^\circ$ to 6$^\circ$ to 2$^\circ$ in the FF scenario.

\subsubsection{Simulation Results}
\label{sec:simres}

We first discuss the appearance of SRP maps for the studied approaches in Sec. \ref{sec:appearanceSRP}, and compare the matrix and map error in Sec. \ref{sec:approxErr} and the localization accuracy in Sec. \ref{sec:locAcc}.

\paragraph{Appearance of SRP Maps}
\label{sec:appearanceSRP}
In an ideal acoustic map, the power in each candidate location would be influenced only by a source in its close vicinity.
The appearance of SRP maps by contrast is dominated by hyperboloids of constant TDOAs \cite{salvati2017exploiting}, with one hyperboloid per microphone pair and (image) source.
The $P$ hyperboloids due to the source intersect at the source location, yielding a large SRP value.
This justifies the commonly used peak picking-based location estimation in (\ref{eq:qshat})--(\ref{eq:imax}).
The intersections of the hyperboloids due to image sources cause a multitude of more or less uniformly distributed local maxima, which can result in localization errors in particular if additional noise is present.
For non-stationary source signals, the strength of individual hyperboloids varies over frames, causing fluctuations in the SRP map.

Fig. \ref{fig:nf_map} shows an NF example of a conv. {FD-}SRP map as well as the corresponding LR-SRP, SLRI-SRP, and SSPI-SRP maps with relative complexity $C_\mathrm{rel} \approx 2\cdot 10^{-2}$. 
The hyperboloid-based appearance described above can be verified in the conv. {FD-}SRP map in Fig. \ref{fig:nf_map}.
We observe that conv. {FD-}SRP correctly localizes the source (i.e. the estimated [\ref{mapExamples_nf_1}] coincides with the true [\ref{mapExamples_nf_0}] source location). 
The LR-SRP and the SLRI-SRP maps are less spatially variant and appear to be defined by only a subset of the hyperboloids apparent in the conv. {FD-}SRP map.
Most of the $P = 6$ hyperboloids due to the source cannot be visually identified anymore and seem to be lost through the low-rank approximation.
As a consequence, the maximum of the SRP maps does not correspond to the source location, resulting in erroneous localization.
The SSPI-SRP map in contrast agrees very well with the conv. {FD-}SRP map, resulting in equally accurate localization.

Fig. \ref{fig:ff_map} shows an FF example of a conv. {FD-}SRP map as well as the corresponding LR-SRP, SLRI-SRP, and SSPI-SRP maps with relative complexity $C_\mathrm{rel} \approx 1 \cdot 10^{-2}$. 
The FF maps are determined by the asymptotic behaviour of the hyperboloids of constant TDOAs, which are transformed to angular coordinates.
Again, we observe that conv. {FD-}SRP and SSPI-SRP localize the source nearly correctly, and that the LR-SRP and the SLRI-SRP maps appear to be defined by only a subset of the hyperboloids apparent in the conv. {FD-}SRP map.
While LR-SRP fails to localize the source, the  SLRI-SRP map preserves a few more features of the conv. {FD-}SRP map and achieves nearly accurate localization, although the maximum of the map is much less pronounced.

The appearance of the LR-SRP, SLRI-SRP, and SSPI-SRP maps can be interpreted as follows.
Due to the nature of low-rank approximations, which approximate matrices by a truncated sum of outer products, LR-SRP and SLRI-SRP introduce additional dependencies between the SRP values across different candidate locations. 
More precisely, we note that the SRP vectors $\mathbf{z}_\mathrm{lr}$ and $\mathbf{z}_\mathrm{slri}$ are forced to lie in the (real part of the) subspace spanned by the columns of $\mathbf{H}_{\mathrm{tall}} \in \mathbb{C}^{J\times R_H}$ and  $\boldsymbol{\Lambda}_{\mathrm{tall}}  \in \mathbb{R}^{J\times R_{\Lambda}}$, respectively, cf. (\ref{eq: z_lr}) and (\ref{eq:z_circlr_computed}).
However, restricting the subspace wherein the SRP vector resides conflicts with the ideal of a spatial map wherein the power in each candidate location would be influenced by a source in its close vicinity only, that is where each source location is represented by a standard basis vector in a $J$-dimensional vector space.
We can therefore argue that the low-rank approximations in LR-SRP and SLRI-SRP {are potentially} useful to the extent that the TDOA-based nature of power computation itself forces dependencies across candidate locations for a given geometry, but are likely to have a deteriorating effect beyond. 
As will be shown in the experiment in Sec. \ref{sec:locAccThreshAperture} and  Sec. \ref{sec:realresults},  low-rank approximations tend to become more efficient with decreasing array apertures and and an increasing number of microphones, respectively. 
SSPI-SRP on the other hand does not restrict the subspace wherein the SRP vector resides.
Moreover, due to the decaying nature of sinc-functions, we can expect many elements of the interpolation matrix $\boldsymbol{\Lambda}$ in (\ref{eq:Lambda_p})--(\ref{eq:Lambda}) to be negligible if the TDOA interval is sufficiently large with respect to the sampling period $T$, which lends it to sparse approximation.

\paragraph{Matrix and Map Error}
\label{sec:approxErr}

Fig. \ref{fig:nf_matMapErr} and \ref{fig:ff_matMapErr} show the matrix error $\varepsilon_{H}$ and the map error $\varepsilon_{z}$ vs. relative complexity $C_{\mathrm{rel}}$ for LR-SRP [\ref{locErr_nf_1}], SI-SRP [\ref{locErr_nf_8}], SLRI-SRP [\ref{locErr_nf_3}], and SSPI-SRP [\ref{locErr_nf_6}] in the NF and FF scenario, respectively.
While $\varepsilon_{H}$ can be computed before run time {and is therefore deterministic},  $\varepsilon_{z}$ is averaged over all 512 source locations, 32 frames, and 16 combinations of $\mathit{SNR}$ and $T_{60}$.
In both the NF and the FF scenario, the characteristics of $\varepsilon_{H}$ and $\varepsilon_{z}$ widely agree across the different algorithms, such that in practice it suffices to compute $\varepsilon_{H}$ in order to judge $\varepsilon_{z}$. 

In both scenarios, LR-SRP can save a bit more or less than one order of magnitude of computational complexity without introducing significant errors.
However, the errors increase rapidly at lower relative complexities.
Unlike LR-SRP, the complexity of SLRI-SRP and SSPI-SRP is restricted to the range between $C_\mathrm{samp}$ and $C_{\mathrm{rel}}$ of SI-SRP, which spans almost three decades in the NF scenario and one decade in the FF scenario.
As compared to  LR-SRP, SLRI-SRP performs similarly in the NF scenario and somewhat better in the FF scenario. 
The better performance of SLRI-SRP in the FF scenario may be due to the more effective SI-SRP approximation as compared to the NF scenario, which leaves room for a less aggressive low-rank approximation of the interpolation matrix.
However, in both scenarios, LR-SRP and SLRI-SRP are significantly outperformed by  SSPI-SRP over a large range of $C_{\mathrm{rel}}$.

\paragraph{{Localization Accuracy for Varying Error Tolerances and Array Aperture}}
\label{sec:locAccThreshAperture}

{Fig. \ref{fig:nf_ff_threshRadius}  shows the localization accuracy $\rho_{\mathrm{s}}$ vs. relative complexity $C_{\mathrm{rel}}$ of conv. TD-SRP [\ref{locErr_nf_10}], LR-SRP [\ref{locErr_nf_1}], SI-SRP [\ref{locErr_nf_8}], SLRI-SRP [\ref{locErr_nf_3}], SSPI-SRP [\ref{locErr_nf_6}], and IR-SRP [\ref{locErr_nf_9}{]} as compared to $\rho_{\mathrm{s}}$ of conv. FD-SRP [\ref{locErr_nf_0}] for different error tolerances $\varepsilon_{\operatorname{th}}$ and diameters $\diameter$ in the NF and FF scenario, respectively, at $\mathit{SNR} = \SI{12}{dB}$ and $T_{60} = \SI{0}{s}$.
In each subplot, the mean of $\rho_{\mathrm{s}}$ is taken over all 512 source locations and 32 frames.
In both scenarios, the absolute performance of the different SRP algorithms largely depends on $\varepsilon_{\operatorname{th}}$ and $\rho_{\mathrm{s}}$ with better performance for larger values, while their relative performance indicates characteristic differences between the algorithms.}

{In the NF scenario, LR-SRP and SLRI-SRP perform very similarly. 
We observe that their perfomance quickly declines for $C_{\mathrm{rel}} < 2\cdot 10^{-1}$ for all choices of $\varepsilon_{\operatorname{th}}$.
For SSPI-SRP, in contrast, we generally observe significantly better performance and a stronger dependency on $\varepsilon_{\operatorname{th}}$, with the point of decline between $C_{\mathrm{rel}} \approx 5\cdot 10^{-3}$ at $\varepsilon_{\operatorname{th}} = \SI{4}{cm}$ and $C_{\mathrm{rel}}  = 1.3\cdot10^{-3}$ at $\varepsilon_{\operatorname{th}} = \SI{16}{cm}$.
These complexities correspond to somewhat less and more than $Q_{\Lambda} = JP$, i.e. using on average one sample per candidate location and microphone pair for interpolation as in conv. TD-SRP.
While the complexity of SSPI-SRP is scalable and allows to trade off complexity and performance, this is not the case for conv. TD-SRP.
Relative to conv. FD-SRP, conv. TD-SRP performs the worse the smaller $\varepsilon_{\operatorname{th}}$, which can be attributed to spatial quantization effects in the SRP map due to the rounding operation in (\ref{eq:TD_SRP}).
The non-exhaustive search approach IR-SRP is particulaly effective in the NF scenario, with the performance close to FD-SRP and the complexity very close to $C_{\mathrm{samp}}$ due to the cost-efficient iterative refinement approach.}

{In the FF scenario, we note that SI-SRP becomes more efficient the smaller the diameter $\diameter$, which is due to $N$ decreasing with microphone distances.
We further observe a key difference between the sparse approaches approaches conv. TD-SRP, SSPI-SRP, and IR-SRP on the one hand and  the low-rank approaches LR-SRP and SLRI-SRP on the other hand.
The smaller the array diameter $\diameter$, the worse perform both TD-SRP and IR-SRP relative to conv. FD-SRP, and the larger the complexity  required by SSPI-SRP to reach the same performance.
This complexity ranges from $C_{\mathrm{rel}} \approx 1.3\cdot 10^{-2}$ at $\diameter=\SI{10}{cm}$ and $C_{\mathrm{rel}} \approx 0.8\cdot 10^{-2}$ at $\diameter=\SI{40}{cm}$, where the letter corresponds to $Q_{\Lambda} = 5JP$, i.e. using on average five samples per candidate location and microphone pair for interpolation. 
The difference in required interpolation accuracy between the NF and the FF scenario may be due to the scenario-dependent relations between candidate location, TDOA, and error metric, cf. (\ref{eq:Deltat_i}) and (\ref{eq:err_s}).
For LR-SRP and SLRI-SRP we observe the opposite relation: the smaller the array diameter $\diameter$, the lower complexities can be reached while maintaining the same complexity as FD-SRP, with SLRI-SRP ($C_{\mathrm{rel}} \approx 1\cdot 10^{-2}$ at $\diameter=\SI{10}{cm}$ and $C_{\mathrm{rel}} \approx 2.3\cdot 10^{-2}$  at $\diameter=\SI{40}{cm}$) outperforming LR-SRP. 
The better performance of SLRI-SRP in the FF scenario may be due to the more effective SI approximation as compared to the NF scenario.
In any of the considered cases, either SSPI-SRP or SLRI-SRP allows for a better trade-off between complexity and performance as compared to TD-SRP and IR-SRP.
At larger array apertures or error tolerances, however, improvements may not be significant.}

\paragraph{Localization Accuracy for Varying Noise and Reverberation Levels}
\label{sec:locAcc}

For the NF and the FF scenario, Fig. \ref{fig:nf_locErr} and Fig. \ref{fig:ff_locErr} respectively illustrate the localization accuracy $\rho_{\mathrm{s}}$ vs. relative complexity $C_{\mathrm{rel}}$ of  {conv. TD-SRP [\ref{locErr_nf_10}]}, LR-SRP [\ref{locErr_nf_1}], SI-SRP [\ref{locErr_nf_8}], SLRI-SRP [\ref{locErr_nf_3}], and SSPI-SRP [\ref{locErr_nf_6}], {and IR-SRP  [}\ref{locErr_nf_9}{]} as compared to $\rho_{\mathrm{s}}$ of conv. {FD-}SRP [\ref{locErr_nf_0}] for different $\mathit{SNR}$ values and reverberation times $T_{60}$.
While the absolute performance of the different SRP algorithms largely depends on the $\mathit{SNR}$ and $T_{60}$, their relative performance exhibits the same trends across different conditions.
As can be seen, the NF scenario is more sensitive to $T_{60}$, cf. also Table \ref{tab:reverbparameters}.

In the NF scenario, we further observe that SSPI-SRP slightly outperforms conv. {FD-}SRP at around $C_{\mathrm{rel}} = 4\cdot 10^{-3}$ (corresponding to $Q_{\Lambda} = 2JP$), indicating that sparse interpolation can even be beneficial to localization performance. 
{As compared to the NF case without reverberation and little noise in Fig. \ref{fig:nf_ff_threshRadius}, we note that the performance of IR-SRP is somewhat more reduced as compared to conv. FD-SRP.
}

\subsection{{Simulations based on Measured RIRs}}
\label{sec:measRIR}

{In addition to simulated RIRs, we perform simulations based on measured RIRs and study the effect of a varying number of microphones.}

\subsubsection{{Simulation Setup}}

{The simulation setup is largely comparable to the NF scenario in Sec. \ref{sec:simsetup}, such that we limit the discussion to the relevant differences.}
{The RIRs with $T_{60} = 0.2\SI{}{s}$ are taken from the MeshRIR database \cite{koyama2021meshrir} and the geometry of the microphone array and the search grid are chosen accordingly.
We use up to $M=13$ microphones lying in a planar grid of $(1 \times 1)\SI{}{m}$ as shown in Fig. \ref{fig:nfreal_geom} at height $\SI{0}{m}$.
More specifically, we perform simulations using $M = \{3,\,5,\,9, \,13\}$ with the microphones selected according to the indeces $m = 1,.., M$ in Fig. \ref{fig:nfreal_geom}, yielding $P = \{3,\,10,\,36, \,78\}$ microphone pairs.
The three-dimensional search grid samples a volume of $(2.4 \times 2.4 \times 0.1) \SI{}{m}$ centered at $[0, 0, -0.1] \SI{}{m}$.
Its spatial resolution is \SI{3.33}{cm}, yielding $J = 21316$ candidate locations.
The 32 sources are regularly distributed around the microphone array with the horizontal coordinates describing a square of $(2 \times 2)\SI{}{m}$ and the height being either $-0.1 \SI{}{m}$ or $0.1\SI{}{m}$. 
As the theoretical TDOAs are the same for positive and negative heights due to the symmetry of the setup, we treat the sources at height $0.1\SI{}{m}$ as if they were located at $-0.1 \SI{}{m}$ in order to use all 32 sources available from the database.
The frame length is $2K = 256$ samples at $\SI{4}{kHz}$, corresponding to \SI{64}{ms}.
For each source location, SRP maps are computed for 128 different frames of microphone data. 
The computational complexity of conv. {FD-}SRP in this scenario is given by $C = \{5.5,\,24.6,\,95.5, \,210.1\}\cdot 10^6$.}

\subsubsection{{Simulation Results}}
\label{sec:realresults}

\paragraph{Localization Accuracy for Varying Number of Microphones}
\label{sec:resultsLocVarMic}

{Fig. \ref{fig:nfreal_locErr} illustrates the localization accuracy $\rho_{\mathrm{s}}$ vs. relative complexity $C_{\mathrm{rel}}$ of {conv. TD-SRP [\ref{locErr_nf_10}]}, LR-SRP [}\ref{locErr_nf_1}{], SI-SRP (single point) [}\ref{locErr_nf_8}{], SLRI-SRP (proposed) [}\ref{locErr_nf_3}{], SSPI-SRP (proposed) [}\ref{locErr_nf_6}{], and IR-SRP (single point) [}\ref{locErr_nf_9}{]  as compared to $\rho_{\mathrm{s}}$ of conv. {FD-}SRP [\ref{locErr_nf_0}] for varying number of microphones $M$  in the NF scenario with measured RIRs. In each subplot, the mean of $\rho_{\mathrm{s}}$ is taken over all 32 source locations and 128 frames.} 

{The peak performance of the different SRP algorithms largely depends on $M$ and begins to saturate for $M>5$.
The curves for  SSPI-SRP exhibits the same trends across different $M$ and drops off below $C_{\mathrm{rel}} = 7\cdot 10^{-3}$ corresponding to using on average a bit more than one sample per candidate location and microphone pair for interpolation. LR-SRP and SLRI-SRP however show a different behavior. While their performance begins to drop at $C_{\mathrm{rel}} <  7\cdot 10^{-2}$ for $M = 3$, it drops only at $C_{\mathrm{rel}} <  3\cdot 10^{-3}$ for $M = 9$ and $M=13$, eventually outperforming SSPI-SRP. This result seems to suggest that above a certain threshold of microphone pairs $P$, additional pairs add redundant information only to the column space of the matrices $\mathbf{H}$ and $\bLambda$, such that low-rank approximations become relatively more efficient. 
We further observe that at a large number of microphones, LR-SRP and SLRI-SRP outperform conv. {FD-}SRP around $C_{\mathrm{rel}} = 5\cdot 10^{-3}$, indicating that low-rank approximations may even be beneficial for source localization.
As in the previous simulations, IR-SRP achieves complexities very close to $C_{\mathrm{samp}}$ at a somewhat reduced performance as compared to conv. {FD-}SRP.}

\definecolor{mycolor1}{rgb}{0.00000,0.44700,0.74100}%
\definecolor{mycolor2}{rgb}{0.85000,0.32500,0.09800}%
\definecolor{mycolor3}{rgb}{0.90000,0.25000,0.25000}%
\definecolor{mycolor4}{rgb}{0.92900,0.69400,0.12500}%
\definecolor{mycolor5}{rgb}{0.25000,0.25000,0.90000}%
\definecolor{mycolor6}{rgb}{0.49400,0.18400,0.55600}%
\definecolor{mycolor7}{rgb}{0.46600,0.67400,0.18800}%
\definecolor{mycolor8}{rgb}{0.30100,0.74500,0.93300}%

\paragraph{{Relative Computation Time versus Relative Complexity}}
\label{sec:resultsComputeTime}

{Fig. \ref{fig:compTime} illustrates the mean relative computation time $D_{\mathrm{rel}}$ vs. relative complexity $C_{\mathrm{rel}}$ of conv. TD-SRP [}\ref{computation_time_1}{], LR-SRP [}\ref{computation_time_2}{], SI-SRP [}\ref{computation_time_3}{], SLRI-SRP (proposed) [}\ref{computation_time_4}{], SSPI-SRP (proposed) [}\ref{computation_time_5}{], and IR-SRP  [}\ref{computation_time_6}{]  for $M = 9$ microphones in the NF scenario with measured RIRs. Here, $C_{\mathrm{rel}}$ of the scalable approaches SL-SRP, SLRI-SRP, and SSPI-SRP is set to $C_{\mathrm{rel}}\approx 5.3\cdot10^{-3}$, corresponding to $C_{\mathrm{rel}}$ of  conv. TD-SRP.  
The gray area indicates the range where $C_{\mathrm{rel}} < C_\mathrm{samp}$ or $D_{\mathrm{rel}} < D_\mathrm{samp}$. The dotted line indicates $D_{\mathrm{rel}} = C_{\mathrm{rel}}$.
The mean of $D_{\mathrm{rel}}$ is taken over all 32 source locations and 128 frames.
{The mean reference duration of conv. FD-SRP is $D = \SI{66.5}{ms}$}.} 

{We observe that $C_{\mathrm{rel}}$ is generally a good predictor of $D_{\mathrm{rel}} $, with all algorithms relatively close to the line $D_{\mathrm{rel}} = C_{\mathrm{rel}}$ in comparison to the range of $C_{\mathrm{rel}}$ (3 orders of magnitude).
The largest deviations with $D_{\mathrm{rel}} \approx 2.5C_{\mathrm{rel}}$ is observed for TD-SRP and SSPI-SRP, which can most likely be attributed to an algorithmic overhead needed for indexing the required TD-GCC samples in TD-SRP (and similarly IR-SRP) and indexing the non-zero elements in $\Lambda_{\operatorname{sp}}$ for SSPI-SRP.
Hardware solutions to reduce this overhead are available in the literature \cite{Zhang2020}.
Further deviations may be due to the abstraction from the computer's hardware in the software architecture of MATLAB, which is  designed as high-level programming language primarily intended for prototyping.
}

\section{Conclusion}
\label{sec:conclusion}

{We proposed two approaches to construct a {fine} SRP map at scalable computational cost, referred to as SLRI-SRP and SSPI-SRP.
Unlike non-exhaustive spatial search approaches, {the proposed approaches rely on a fine grid of candidate locations}. The proposed approaches are based on a decomposition of the {conv. FD-}SRP matrix into a sampling and an interpolation matrix and optimal low-rank or sparse approximations of the latter.
In simulations, the proposed approaches are compared to {conv. FD-SRP, conv. TD-SRP,} SI-SRP, LR-SRP in \cite{grondin2019svd} and the non-exhaustive spatial search approach IR-SRP in \cite{marti13} in {various} speech source localization scenarios including large and small array sizes.
The results indicate that SSPI-SRP performs better if large array apertures are used, while SLRI-SRP performs better at small array apertures or a large number of microphones. 
{In comparison to conv. FD-SRP,} two to three orders of magnitude of complexity reduction can achieved, {often times enabling a more favourable complexity-performance trade-off as compared to conv. TD-SRP.} 
The results indicate that among the fine SRP mapping approaches, SSPI-SRP outperforms both SLRI-SRP and LR-SRP over a wide complexity range 
{if a low number of microphones and large array apertures are used, while SLRI-SRP performs better for small array apertures, and both SLRI-SRP and LR-SRP perform better at a larger number of microphones.}
{In comparison to conv. FD-SRP, two to three orders of magnitude of complexity reduction can achieved, often times yielding a more favourable trade-off between complexity and performance in comparison to conv. TD-SRP.} 
A further exploration of the dependency of the proposed approaches on the geometry of the array and the search grid as well as the relation between the number of microphones and the performance of the low-rank approaches SLRI-SRP and LR-SRP is subject to further research.   
Further, potential applications of low-rank decompositions to the problem of microphone-pair subset selection will be investigated.}

{\appendix[Two-sided Matrix-Vector Formulations]

In the Appendix, we introduce two-sided matrix-vector formulations.
For the two-sided frequency bin index $k \in \{-K+1,\dots,K\}$, let the index ${k}_{\mathrm{mod}}$ be defined by
\begin{align}
{k}_{\mathrm{mod}} &= \operatorname{mod}[{k},2K]+1,
\end{align}
yielding ${k}_{\mathrm{mod}} \in \{1,\dots,2K\}$. 
As two-sided counterparts  of ${\bpsi}_p$, ${\mathbf{H}}_p$, and ${\mathbf{S}}_p$ in (\ref{eq:psi_p_single}), (\ref{eq:H_p_single}), and (\ref{eq:S_p_single}), respectively,
we define $\bar{\bpsi}_p \in \mathbb{C}^{2K}$, $\bar{\mathbf{H}}_p \in \mathbb{C}^{J \times 2K}$, and $\bar{\mathbf{S}}_p \in \mathbb{C}^{N_p \times 2K}$ as
\begin{align}
[\bar{\bpsi}_p]_{{k}_{\mathrm{mod}}} &= \psi_{p}(\omega_{{k}}),\label{eq:psi_p_two}\\
[\bar{\mathbf{H}}_p]_{i,{k}_{\mathrm{mod}}} &= e^{j\omega_{{k}}\Delta t_{p}(i)},\label{eq:H_p_two}\\
[\bar{\mathbf{S}}_p]_{{n}_{\mathrm{mod}},{k}_{\mathrm{mod}}} &= e^{j  \pi {k} n/K}, \label{eq:S_p_two}
\end{align}
Stacking over microphone pairs, as the two-sided counterparts of ${\bpsi}$, ${\mathbf{H}}$, and ${\mathbf{S}}$ in (\ref{eq:psi_single}), (\ref{eq:H_single}), and (\ref{eq:S_single}), we further define 
$\bar{\bpsi} \in \mathbb{C}^{2PK}$, $\bar{\mathbf{H}} \in \mathbb{C}^{J \times 2PK}$, and $\bar{\mathbf{S}} \in \mathbb{C}^{PN \times 2PK}$ as
\begin{align}
\bar{\bpsi} &= \begin{pmatrix}
\bar{\bpsi}^\transp_{1} & \cdots & \bar{\bpsi}^\transp_{P} 
\end{pmatrix}^\transp, \label{eq:psi_two}\\
\bar{\mathbf{H}} &= \begin{pmatrix}
\bar{\mathbf{H}}_1 & \cdots &\bar{\mathbf{H}}_P 
\end{pmatrix}, \label{eq:H_two}\\
\bar{\mathbf{S}} &= 
\operatorname{blkdiag} \bigl[
\bar{\mathbf{S}}_1,\, \dots,\, \bar{\mathbf{S}}_P
\bigr],\label{eq:S_two}
\end{align}
With these two-sided definitions, $\mathbf{z}$ in (\ref{eq:z_computed}) can alternatively be expressed by
\begin{align}
\mathbf{z} &= \sum_{p=1}^P \boldsymbol{\xi}_{\Delta t_p}\nonumber\\
&= \sum_{p=1}^P \bar{\mathbf{H}}_p \bar{\bpsi}_p\nonumber\\
&= \bar{\mathbf{H}} \bar{\bpsi}. \label{eq:z_computed_two}
\end{align} 
Similarly, we can alternatively write $\boldsymbol{\xi}_p = \bar{\mathbf{S}}_p \bar{\bpsi}_p$ and accordingly express $\boldsymbol{\xi}$ in (\ref{eq:xi_computed}) as
\begin{align}
\boldsymbol{\xi} &= \bar{\mathbf{S}} \bar{\bpsi}. \label{eq:xi_computed_two}
\end{align} 
The correspondence between the two-sided formulations (\ref{eq:z_computed_two}) and (\ref{eq:xi_computed_two}) and the singe-sided formulations (\ref{eq:z_computed}) and (\ref{eq:xi_computed}) is exact if $\psi_{p}(0) = 0$ and $\psi_{p}(\pm\omega_{\mathrm{b}}) = 0$, cf. Sec. \ref{sec:discreteFreq}.
Note that it can easily be verified that the rows of $\bar{\mathbf{S}}_p$ correspond to the rows of the $2K$-iDFT matrix at indeces $n^\prime_{\mathrm{mod}}$, with $n^\prime_{\mathrm{mod}}$ as defined in (\ref{eq:n_prime}).
This allows the use of the iFFT in (\ref{eq:xi_p_ifft})--(\ref{eq:xi_p_from_ifft}).
From this relation to the $2K$-iDFT matrix, we further observe that the rows of $\bar{\mathbf{S}}_p$ are orthogonal such that
\begin{align}
\bar{\mathbf{S}} \bar{\mathbf{S}} ^\herm = 2K\mathbf{I},
\end{align}
which is exploited in (\ref{eq:cost_simplified}).

 }

\section*{Acknowledgments}

This research work was carried out at the ESAT Laboratory of KU Leuven, in the frame of KU Leuven internal funds C14/21/075 "A holistic approach to the design of integrated and distributed digital signal processing algorithms for audio and speech communication devices", C24/16/019 "Distributed Digital Signal Processing for Ad-hoc Wireless Local Area Audio Networking" and VES/19/004, and FWO Large-scale research infrastructure "The Library of Voices - Unlocking the Alamire Foundation’s Music Heritage Resources Collection through Visual and Sound Technology" (I013218N). The research leading to these results has received funding from the European Research Council under the European Union's Horizon 2020 research and innovation program / ERC Consolidator Grant: SONORA (no. 773268). This paper reflects only the authors' views and the Union is not liable for any use that may be made of the contained information.

\bibliographystyle{IEEEtran}
\bibliography{}

\newpage

\vfill

\end{document}